\documentclass[10pt]{article}
\usepackage[onecolumn]{mnjen}

\usepackage{amstex}
\usepackage{epsfig}
\usepackage{amssymb}
\usepackage{eepic}

\newcommand{\be}{\begin{equation}}
	{\begin{equation}}%
{\end{equation}}

\def\modenergy{\left| \frac{\beta \Util^2 +\beta -2}{2\beta} \right|  } 
\def\betagammabig{  \frac{1}{\beta} + \frac{\gamma}{\beta} - \frac{\gamma}{2} }
\def\betagammasmall{  1/\beta + \gamma/\beta - \gamma/2 }

\def\auxint{ \mathcal{J} }
\def\flmint{ \mathit{J} }

\def\fr#1#2{\textstyle {#1\over #2}\displaystyle}

\def\d2rdt2{{ {{d^2}R}\over {dt^2} }}
\def\half{{ {1 \over 2} }}
\def\RH{ R_{\text{H}} }
\def\Rtil{ \tilde{R} }
\def\RtilH{ \tilde{R}_{\text{H}} }
\def\Rtilc{ \tilde{R}_{\text{c}} }
\def\Rtilmin{ \tilde{R}_{\min} }
\def\Rtilmax{ \tilde{R}_{\max} }
\def\Util{ \tilde{U} }
\def\Ltil{ \tilde{L}_z }
\def\Ltilc{ \tilde{L}_{\text{c}} }
\def\Lc{ L_{\text{c}} }

\def\Etil{ \tilde{E} }
\def\Htil{ \tilde{H} }

\def\Ytil{ \tilde{Y} }
\def\Xtil{ \tilde{X} }
\def\xtil{ \tilde{x} }

\def\Atil{ \tilde{A} }
\def\Atilimp{ \tilde{A}_{\text{imp}} }

\def\Aimp{ A_{\text{imp}}}
\def\Ares{ A_{\text{res}}}

\def\Sigimp{ \Sigma_{\text{imp}} }
\def\Sigeq{ \Sigma_{\text{eq}} }
\def\Sigres{ \Sigma_{\text{res}} }
\def\Sigmap{ \Sigma_{\text{p}} }
\def\fsing{ f_{\text{s}} }
\def\fcutout{ f_{\text{cutout}} }
\def\fimp{ f_{\text{imp}} }
\def\psimp{ \psi_{\text{imp}} }
\def\vcirc{ v_{\text{circ}} }

\def\sigutil{ \tilde{\sigma}_u }

\def\kaptil{ \tilde{\kappa} }
\def\Omegap{ \Omega_{\text{p}} }
\def\Omtil{ \tilde{\Omega} }

\def\omtil{ \tilde{\omega} }

\def\LtilcNo{ \tilde{L}_c^{M_\beta} }

\def\LNi{ L_z^{N_\beta} }
\def\LNo{ L_z^{M_\beta} }
\def\Sm{ \mathcal{S}_m }
\def\Smtil{ \tilde{\mathcal{S}}_m }
\def\Qlm{ Q_{lm}}
\def\Flm{ F_{lm}}
\def\lkapmom{ l\tilde{\kappa} + m\tilde{\Omega} }
\def\twopm{\frac{2+\beta}{2-\beta}}
\def\mtwopm{\frac{2+\beta}{\beta-2}}
\def\twomp{\frac{2-\beta}{2+\beta}}
\def\modterm{ \left| \frac%
	{  2 + 2\gamma - \beta \gamma }%
	{ \beta\Util^2 + \beta  - 2   } \right| }
\def\exprei{ e^{\frac{2j-1}{N} \pi \eta } }
\def\expreid{ e^{ \frac{2j-1}{N} \pi \hat{\eta}} }
\def\expimi{ e^{ \frac{2j-1}{N} i\pi } }
\def\expmNiNo{ e^{ -(2k-1) \frac{N}{M} i\pi } }
\def\expimo{ \Ltilc^{\twopm} e^ { \frac{2k-1}{M} i\pi  } }
\def\imbracki{ \left[ \lkapmom - \omtil \expimi \right] }
\def\imbracko{ \left[ \lkapmom - \omtil \expimo \right] }

\title{Stability of Power-Law Disks I -- The Fredholm Integral
Equation}
\author[N. W. Evans and J. C. A. Read]{N. W. Evans and J. C. A. Read 
\\Theoretical Physics, Department of Physics, 1 Keble Rd, Oxford, OX1 3NP} 

\pubyear{1997}
 
\begin{document}

\maketitle

\begin{abstract}
\noindent
The power-law disks are a family of infinitesimally thin, axisymmetric
stellar disks of infinite extent. The rotation curve can be rising,
falling or flat. The self-consistent power-law disks are scale-free,
so that all physical quantities vary as a power of radius. They
possess simple equilibrium distribution functions depending on the two
classical integrals, energy and angular momentum.  While maintaining
the scale-free equilibrium force law, the power-law disks can be
transformed into cut-out disks by preventing stars close to the origin
(and sometimes also at large radii) from participating in any
disturbance.

This paper derives the homogeneous Fredholm integral equation for the
in-plane normal modes in the self-consistent and the cut-out power-law
disks.  This is done by linearising the collisionless Boltzmann
equation to find the response density corresponding to any imposed
density and potential.  The normal modes -- that is, the
self-consistent modes of oscillation -- are found by requiring the
imposed density to equal the response density. In practice, this
scheme is implemented in Fourier space, by decomposing both imposed
and response densities in logarithmic spirals. The Fredholm integral
equation then relates the transform of the imposed density to the
transform of the response density. Numerical strategies to solve the
integral equation and to isolate the growth rates and the pattern
speeds of the normal modes are discussed.

\end{abstract}
 


\section{Introduction}

\noindent
This paper begins an investigation into the large-scale, global,
linear modes of a family of horizontally hot, vertically cold,
idealised disk galaxies with rising, falling or flat rotation curves.
Here, we collect all the mathematical and numerical details needed for
our study. A companion paper completes the investigation by presenting
results on the global spiral modes, together with the astrophysical
implications.  The focus is exclusively on the in-plane instabilities,
such as the modes causing bar-like or lop-sided distortions.  Of
course, our hot, razor-thin models are also unstable to bending modes
\cite{MerrSell:1994}. These are ignored because they almost 
certainly can be eliminated by giving the disks a modest thickness.

In our study, we have the good fortune to be able to follow a
magisterial earlier analysis carried out by
Zang~\shortcite{Zang:1976}.  This remarkable Ph. D. thesis was
supervised by Alar Toomre and also benefited from a number of
ingenious suggestions provided by Agris Kalnajs.  What Zang did in the
mid-seventies was to carry out the first complete, global, stability
analysis for a family of differentially rotating, stellar dynamical
disks. The object of his study was an infinitesimally thin disk in
which the circular velocity is completely flat. The special case in
which the stars move on circular orbits is known as the
~\longcite{Mestel:1963} disk. Their hot stellar dynamical counterparts
are usually called the Toomre-Zang disks. The global spiral modes in
this model were described extensively by Zang.

The power-law disks are infinitesimally thin disks in which the
circular velocity varies as a power of cylindrical polar radius, viz.,
$\vcirc \propto R^{-\beta/2}$. They are amongst the simplest stellar
dynamical disks known -- the distributions of stellar velocities that
build the models are given by ~\longcite{Evans:1994}. Their rotation
curves can be rising or falling.  The self-similarity of these disks
simplifies the analysis considerably. It enables much of the global
normal mode calculation to be performed exactly.  The model examined
in Zang's~\shortcite{Zang:1976} thesis is the particular case with a
completely flat rotation curve ($\beta =0$). The perfectly
self-similar disks are somewhat special. To supplement our analysis,
we have also concentrated on a modified version of the pure power-law
disks, in which the central regions of the disks are immobilised or
cut-out. This was achieved by reducing the fraction of active stars --
those able to respond to any disturbance -- from unity at moderate
radii to zero at the centre. In order to aid comparison of our results
with numerical simulations, especially those of
Earn~\shortcite{Earn:1993}, we have also tapered the fraction of
active stars to zero at large radii. The immobile components still
contribute to the potential experienced by the stars.  The material in
Zang's~\shortcite{Zang:1976} thesis was never completely published,
although some of these and various subsequent calculations are
reported in Toomre~\shortcite{Toomre:1977,Toomre:1981}. So, we urge
the reader to remember that our contribution consists {\em only} in
generalising the analysis of Zang from the disk with a completely flat
rotation curve to the entire family of power-law disks. This is not
trivial, as the limit of a flat rotation curve is a singular
one. Nonetheless, our job has been very considerably eased by being
able to lean on the sturdy support of Zang and his two implicit
co-workers. We use the notation Z followed by an equation number as a
convenient shorthand for reference to Zang's~\shortcite{Zang:1976}
thesis.

The aim of this paper is to derive the homogeneous, singular, Fredholm
integral equation for the normal modes in the self-consistent and the
cut-out power-law disks. First, the equilibrium properties of the
models are introduced in Section 2. The collisionless Boltzmann
equation is linearised to derive the Fredholm integral equation in
Section 3. Section 4 summarises the strategies for its computational
solution. This paper describes only methods. The following paper in
this issue of {\it Monthly Notices} presents the numerical results and
astrophysical consequences.

\section{The Equilibrium disks}\label{sec:EqbmChap}

\noindent
This section begins with a brief introduction to the power-law
disks. Section 2.2 introduces new variables -- the home radius and the
eccentric velocity -- which can be used to characterise the stellar
orbits according to their size and shape. The frequencies and the
periods of orbits are derived in the next section. The equilibrium
distribution functions of stellar velocities of the self-consistent
power-law disks are presented in Section 2.4. The concluding section
explains how to ``cut out'' the centre of the disk and to taper the
disk at large radii. Both these require changes to the distribution
function.

\subsection{The Potential-Density Pair}
 
\noindent
The equilibrium density of the power-law disks is
\begin{equation}         
        \Sigeq  = \Sigma_0 \left( R_0 \over R \right) ^{1+\beta} 
	\label{eq:surfacedensity}.
\end{equation}
The self-consistent potential in the plane of the disk is 
~\cite{SE:1987,LKLB:1991,Evans:1994}
\begin{equation} 
     \psi (R) =   \frac{v_\beta^2}{\beta} \left( R_0 \over R
\right)^\beta \label{eq:potential},
\end{equation} 
where the reference velocity $v_\beta$ is defined as
\be 
        v_\beta^2 = 2 \pi G \Sigma_0 R_0  
         {{ \Gamma \left[ \half \left( 1-\beta \right)  \right]   
         \Gamma \left[ \half \left( 2+\beta \right)  \right] }  
        \over 
         { \Gamma \left[ \half \left( 1+\beta \right)  \right]   
         \Gamma \left[ \half \left( 2-\beta \right) \right] } } 
	\label{eq:vbeta} .
\end{equation} 
Clearly~\eqref{eq:potential} fails for $\beta=0$. In this case, we 
have (\citenobrack{Mestel:1963})
\begin{equation} 
     \psi (R) = -v_0^2 \ln \left( {R \over R_0}\right) .
\end{equation} 
Whenever an expression fails at $\beta=0$, the corresponding result
for the Toomre-Zang disk should be used. It can often be derived by
taking the limit $\beta \rightarrow 0$ with l'H\^{o}pital's
theorem. The circular velocity is
\be 
         \vcirc^2 = \left( R_0 \over R \right) ^\beta v_\beta^2  .
	\label{eq:circvel}
\end{equation} 
Thus, the reference velocity $v_\beta$ is the circular velocity at the
reference radius $R_0$. Disks with $\beta>0$ have falling rotation
curves, whereas disks with $\beta <0$ have rising rotation curves.
Disk galaxies typically have flattish rotation curves
(e.g. \citenobrack{Rubin:1978,MnB}) -- sometimes slowly rising,
sometimes slowly falling at large radii~\cite{CasvG:1991}.  The total
mass of the disk is infinite for all $\beta$ in the range $-1 \le
\beta < 1$. The mass enclosed within a radius $R$ is
\be
	M(R) = \frac{2\pi\Sigma_0 R_0^2}{1-\beta} 
	\left( \frac{R}{R_0} \right)^{1-\beta}.
	\label{eq:MassL}
\end{equation}
%


\subsection{The Home Radius and the Eccentric Velocity} 

\noindent
The Lagrangian for stars orbiting in a general power-law disk is
\begin{equation} 
\mathcal{L} =  \half \dot{R}^2  + \half R^2 \dot{\theta}^2 + { v_\beta^2 \over \beta} \left( 
{ R_0 \over R} \right) ^\beta \label{eq:Lagrangian} .
\end{equation} 
There are two isolating integrals of motion, namely the energy $E$ and
the angular momentum $L_z$.  In terms of the radial velocity $u =
\dot{R}$ and the tangential velocity $v = R \dot{\theta}$, the
integrals are
\begin{xalignat}{2}
        L_z & = Rv, 
& \quad
        E & = \half \left( u^2 + v^2 \right) - {v_\beta^2 \over \beta} \left( { R_0 \over R} \right) 
^\beta \label{eq:energyR} .
\end{xalignat}
Following Zang~\shortcite{Zang:1976}, let us define the {\em home radius}
$\RH$ to be the radius at which the tangential velocity is equal to
the circular velocity. By conservation of angular momentum, we have
\be 
\RH = R_0 \left( { L_z \over {v_\beta R_0} } \right)^{2 \over {2-\beta}}.
\label{eq:homeradius} 
\end{equation} 
Again following Zang~\shortcite{Zang:1976}, the {\em eccentric
velocity} $U$ is defined as the maximum radial speed reached during an
orbit. $U$ is thus positive by definition. It is easy to show that the
eccentric velocity is attained at the home radius
\eqref{eq:homeradius} and that its value is given by
\begin{equation} 
        U^2 = 2E + \left( {2 \over \beta } - 1 \right)  
           \left( { {v_\beta^2 R_0^\beta} \over L_z^{\beta} } \right) ^ { 2 \over {2-\beta} } 
\label{eq:eccvel2}.
\end{equation} 
A similar derivation using the potential of the Mestel disk gives the
result already well-known to Zang (Z2.29)
\begin{equation} 
        U^2 = 2E - v_0^2 \left( 1 + 2\ln{\frac{L_z}{v_0 R_0}} \right) .
\end{equation} 
We shall find it convenient to work in dimensionless co-ordinates. We
define the following dimensionless integrals of motion:
\begin{xalignat}{4}
\Util^2  & = {U^2 \over v_\beta^2} \left( {\RH \over R_0} \right) ^\beta, 
& \quad 
\RtilH & = {\RH \over R_0}, & \quad
\Etil & = {E \over v_\beta^2} \left( {\RH \over R_0} \right) ^\beta   ,
&\quad  
\Ltil & = { L_z \over {v_\beta R_0} } .
\label{eq:ELnondim}
\end{xalignat} 
Likewise, dimensionless radial and tangential velocities, radius and
time coordinates can be defined as:
\begin{xalignat}{4}
	\tilde{u}^2 & =  
{u^2 \over v_\beta^2} \left( {\RH \over R_0} \right) ^\beta  ,
& \quad
	\tilde{v}^2 & = {v^2 \over v_\beta^2} 
        \left( {\RH \over R_0} \right) ^\beta ,
& \quad
	\Rtil & = {R \over \RH} ,
& \quad 
\tilde{t} & = {{v_\beta} \over {\RH}}  
\left( {{R_0} \over {\RH}} \right)^{\beta/2} t .
\label{eq:utilvtil}
\end{xalignat} 
The scaled energy~\eqref{eq:energyR}, home radius~\eqref{eq:homeradius} 
and eccentric velocity~\eqref{eq:eccvel2} are given by
\begin{xalignat}{3}
	\Etil & = \half \left( \tilde{u}^2 + \tilde{v}^2 \right) - {1 \over {\beta \Rtil^\beta} } ,
&\quad
	\RtilH & = \Ltil^{2 \over {2 - \beta}} ,
& \quad
	\Util^2 & =  2 \Etil -1 + { 2 \over \beta}. 
	\label{eq:Util} 
\end{xalignat} 
We will also need expressions for the scaled radial and tangential
velocities:
\begin{xalignat}{2} 
	\tilde{u}^2 & = \Util^2 + 1 - \Rtil^{-2} + {2 \over \beta} \left( {\Rtil^{-\beta} -1 } 
\right) \label{eq:dimradvel} ,
& \quad
	\tilde{v} & = {1 \over \Rtil}\label{eq:dimtangvel}. 
\end{xalignat}
In terms of the scaled coordinates, the equations of motion derived
from the Lagrangian become
\begin{xalignat}{2}
	{{d^2 \Rtil} \over {d \tilde{t}^2}} 
	& = {1 \over {\Rtil^3}} - {1 \over {\Rtil^{1+\beta}} } ,
&\quad
	\frac{d\theta}{d \tilde{t}} = \frac{1}{\Rtil^2}
\label{eq:dimeqmot}.
\end{xalignat}
Imagine solving the equations of motions with starting positions and
velocities. If two stars have the same $\tilde{u}$ and $\tilde{v}$,
then their orbits have the same shape, although the size of the orbit
is proportional to $\RH$. This leads us to an intuitive understanding
of the eccentric velocity and the home radius. Stellar orbits with the
same dimensionless eccentric velocity have the same shape. The
dimensionless home radius fixes the overall size of the orbit.

The star's orbit in the equilibrium disk is limited by its isolating
integrals of motion. In terms of the dimensionless co-ordinates, the
turning-points of the orbit occur when
\be \label{eq:extrema} 
	\tilde{u}^2 = \Util^2 + 1 - \Rtil^{-2} + {2 \over \beta}  
      \left( \Rtil^{-\beta} - 1 \right) = 0.
\end{equation} 
For given $\Util$ and $\beta$, there are two solutions, $\Rtilmin $
corresponding to pericentre, and $\Rtilmax$ corresponding to
apocentre.  These are easy to find numerically by the Newton-Raphson
technique.  The star thus moves within an annulus of the disk. For
certain values of the eccentric velocity, the orbit closes, and the
star then traverses a one-dimensional manifold within the annulus; in
general, the orbit does not close, and the star eventually passes
arbitrarily close to every region of the annulus.


\subsection{Periods and Frequencies}\label{sec:perfreq} 
Let us start by defining a useful auxiliary integral 
\begin{equation} 
	\auxint_n (\Util) = 2 \int_{\Rtilmin(\Util)}^{\Rtilmax(\Util)}  
{ 
{d\Rtil} 
\over 
{\Rtil^n 
  \left( \Util^2 + 1 - \Rtil^{-2}  
+ {2 \over \beta} \left( \Rtil^{-\beta} - 1 \right) \right) ^\half  
} 
}\label{eq:genintI} ,
\end{equation} 
where $\Rtilmin$ and $\Rtilmax$ are the solutions of~\eqref{eq:extrema}. 
For the $\beta=0$ case, this becomes (Z2.39)
\begin{equation} 
	\auxint_n (\Util) = 2 \int_{\Rtilmin(\Util)}^{\Rtilmax(\Util)}  
{ 
{d\Rtil} 
\over 
{\Rtil^n 
  \left( \Util^2 + 1 - \Rtil^{-2} -2 \ln \Rtil  \right) ^\half  
} 
} 
\label{eq:genintIbzero} .
\end{equation} 
To evaluate $\auxint_n$ numerically, we remove the singularities at
either end of the integrand by transforming to a variable $\theta$. We
define $\Rtil = m + a \sin\theta$, where $m$ is the midpoint of the
radial motion, $m = \half ( \Rtilmin + \Rtilmax )$ and $a$ is its
amplitude, $a = \half ( \Rtilmin - \Rtilmax )$. The integration can
then be carried out using the midpoint method~\cite[chap. 4]{NumRec}.

\begin{figure}
\begin{center}
\epsfig{file=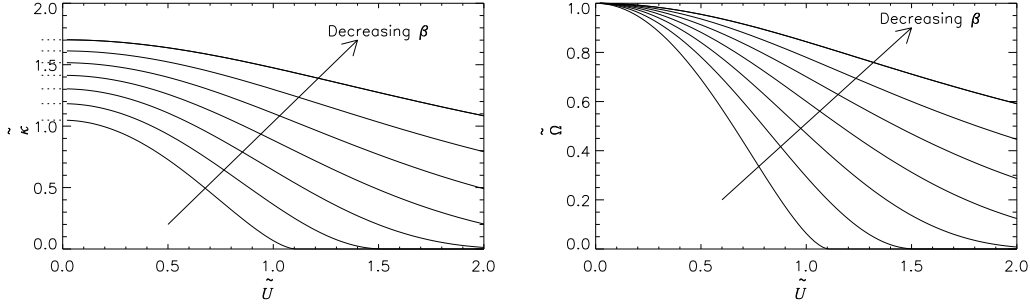,width=0.8\textwidth}
\caption[Radial and angular frequencies against
eccentric velocity] {The radial frequency $\kaptil$ and angular
frequency $\Omtil$ plotted against eccentric velocity $\Util$, for
$\beta$ = -0.9, -0.6, -0.3, 0, +0.3, +0.6, +0.9. In the left-hand
plot, the dotted lines on the vertical axis mark the epicyclic limit
of $\sqrt{2-\beta}$. As $\Util \rightarrow \infty$, the frequencies 
$\kaptil$ and $\Omtil$ remain positive but become vanishingly small.
\label{fig:kapOm}}
\end{center}
\end{figure}

The radial period $T$ is the time taken for the star to travel between
two successive pericentres: i.e., to move out and in again. The
corresponding radial frequency is defined by $\kappa = 2\pi/T$. Using
the symmetry of the orbit, we have
\be 
	T = 2 \int_{R=R_{\min}}^{R=R_{\max}} dt = 2 \int_{R_{\min}}^{R_{\max}} 
{dR \over \dot{R} } 
	= { {2 \RH } \over v_\beta} \left( {\RH \over R_0} \right) ^ {\beta \over 2} 		     
	\int_{\Rtilmin}^{\Rtilmax} {d\Rtil \over \tilde{u} }.
\end{equation} 
Using the expression for the radial velocity given
in~\eqref{eq:dimradvel}, along with the definition~\eqref{eq:genintI},
we find that the radial period and frequency are
\begin{xalignat}{2} 
	T &= {\RH \over v_\beta} \left( {\RH \over R_0} \right) ^ {\beta \over 2} \auxint_0 ,
&\quad
	\kappa = { {2\pi} \over T} 
	= { v_\beta \over \RH } \left( { R_0 \over \RH }
         \right) ^ {\beta \over 2} {{2\pi} \over \auxint_0}.
\end{xalignat}
We shall find it useful to define the dimensionless radial frequency $\kaptil$:
\be 
\kaptil = { \RH \over v_\beta } \left( { \RH \over R_0 } \right) ^ {\beta \over 2} \kappa 
	= {{2\pi} \over \auxint_0}.  
\label{eq:kaptil}
\end{equation} 
The angular period $\Theta$ is defined to be the angle through which a
star moves during the time taken to complete one radial oscillation:
\be 
	\Theta = \int_{t=0}^{t=T} d\theta  
	= 2 \int_{R_{\min}}^{R_{\max}} {{\dot{\theta} } \over \dot{R} } dR  
	= 2 \int_{R_{\min}}^{R_{\max}} { {v dR} \over {u R} } 
	= 2 \int_{\Rtilmin}^{\Rtilmax}
 { 
 {\tilde{v} d\Rtil} 
 \over 
 {\tilde{u} \Rtil } 
 } 
	= \auxint_2.
\end{equation}
The angular frequency $\Omega$ is defined to be the mean angular speed
of the star: $\Omega = \Theta / T$. We shall commonly use the
dimensionless angular frequency $\Omtil$, where
\be 
	\Omtil = { \RH \over v_\beta } \left( { \RH \over R_0 } \right) ^ {\beta \over 2} 
\Omega 
	= {\auxint_2 \over \auxint_0} .
	\label{eq:Omtil}
\end{equation} 
The dependence of $\kaptil$ and $\Omtil$ on $\Util$ and $\beta$ is
shown in Fig.~\ref{fig:kapOm}.
In the limit $\Util \rightarrow 0$, the dimensionless epicyclic and
circular frequencies are
\begin{xalignat}{2}
	\kaptil_0 \equiv \lim_{ \Util\rightarrow 0 } \kaptil  &= \sqrt{2 - \beta} ,
& \quad
	\Omtil_0 \equiv \lim_{ \Util\rightarrow 0 } \Omtil & = 1 .
	\label{eq:kaptil0Omtil0}
\end{xalignat}
In the limit $\Util \rightarrow \infty$, these frequencies tend to
zero from above.  As expected, the angular frequency is that of a star
in a circular orbit at $\RH$, namely $\Omega_0 = \vcirc / \RH$. At low
eccentric velocity, there is superimposed on this circular motion a
small radial oscillation with frequency $\kappa_0 = \sqrt{2-\beta}\,
\vcirc/ \RH$.  This is of course just the epicyclic approximation for
near-circular orbits~\cite{BnT}.

\subsection{The Structure of Phase Space\label{sec:StructVel}}	
\noindent
The distribution function of the stars can depend on positions and
velocities only through the isolating integrals of motion, according
to Jeans'~\shortcite{Jeans:1919} theorem. Even quite simple disks can
possess rather complicated distribution
functions~\cite{EvCol:1993}. The power-law disks are attractive
candidates for a stability analysis because they possess a rich family
of simple self-consistent distribution functions $\fsing$ found by
Evans~\cite*{Evans:1994}. These are built from powers of energy and
angular momentum, and normalised so that the integral of the
distribution function over all velocities recovers the surface
density. They are:
\begin{xalignat}{2}
	\fsing(E,L_z) & =  \tilde{C}
	L_z^{\gamma} | E | ^{\betagammasmall} 
	\label{eq:fsing},
& \quad	
 \text{where}  \hspace{1cm}
	\tilde{C} & = {
{ C_{\beta \gamma} \Sigma_0 | \beta | ^{1+{1 \over \beta} + {\gamma \over \beta} } } 
\over 
{2^{\gamma / 2} \sqrt{\pi} R_0^\gamma 
v_\beta^{2 \left(1+{1 \over \beta} + {\gamma \over \beta} \right)} } 
} 
	\label{eq:Ctilde};  
\end{xalignat}
with
\begin{xalignat}{3}
\beta & > 0: 
&\quad
	 C_{\beta \gamma} & = { {\Gamma \left[ 2+{1 \over \beta} + {\gamma \over \beta} \right]} 
\over 
{\Gamma \left[ \half (\gamma + 1) \right] \Gamma \left[ 1 + {1 \over \beta} + {\gamma \over \beta} - {\gamma \over 2} \right] } },
& \quad
\text{and} \hspace{1cm} \gamma &> -1; 
\\
\beta &< 0:
& \quad
 	C_{\beta \gamma} & = { 
{\Gamma \left[ {\gamma \over 2} - {1 \over \beta} - {\gamma \over \beta} \right]} 
\over 
{\Gamma \left[ \half (\gamma + 1) \right] 
\Gamma \left[ -1 - {1 \over \beta} - {\gamma \over \beta} \right] } 
} ,
&\quad
\text{and} \hspace{1cm} \gamma & > -\beta -1 .
\end{xalignat}
Note that these formulae differ by a factor of two from those given in
Evans'~\cite*{Evans:1994} paper. This is because Evans' results assume
a bi-directional disk, where the stars rotate in both senses. In this
paper, we consider only uni-directional disks. Analogous distribution
functions for the Toomre-Zang disk have been known for a long
time~\cite[chap. 4]{BK:1976,Zang:1976,Toomre:1977,BnT}
\begin{xalignat}{3}
	\fsing(E,L_z) & = \tilde{C} L_z^\gamma \exp \left( -(\gamma+1){E \over {v_0^2} }\right),
\label{eq:fsingMestel}
& 
	\text{ where   } 
	\tilde{C} & = { {C_{0 \gamma} \Sigma_0}  \over {2^{\gamma / 2} \sqrt{\pi} R_0^\gamma v_0^{\gamma+2}} },
	\label{eq:CtildeMestel}
& 
	C_{0 \gamma} & = { { \left( \gamma + 1 \right)^{1+\gamma / 2} } \over { \Gamma \left[ \half (\gamma + 1) \right]  } }.
\end{xalignat}
In these formulae, $\gamma$ is a constant anisotropy parameter. Its
physical meaning will shortly become apparent on examining the
dynamical quantities derived from the distribution function.

The mean streaming velocity, or the stellar rotation curve, is
\begin{xalignat}{2}
	\langle v \rangle &=	
	 v_\beta  \sqrt{\frac{2}{\beta}}
	\frac
	{\Gamma \left[ 1+\frac{\gamma}{2}\right]  }
	{\Gamma \left[ \frac{1}{2} + \frac{\gamma}{2} \right] }
	\frac
	{ \Gamma \left[ 2+\frac{1}{\beta}+\frac{\gamma}{\beta} \right]}
	{\Gamma \left[ \frac{5}{2} +\frac{1}{\beta}+\frac{\gamma}{\beta}\right]} 	
	\left( \frac{R_0}{R} \right) ^{\beta/2},
& \quad \beta > 0,\\
	\langle v \rangle &= v_\beta \sqrt{\frac{2}{-\beta}}
	\frac
	{\Gamma \left[ 1+\frac{\gamma}{2} \right] 
	\Gamma \left[ -\frac{3}{2}-\frac{1}{\beta}-\frac{\gamma}{\beta} \right] }
	{\Gamma \left[ \frac{1}{2}+\frac{\gamma}{2} \right] 
	\Gamma \left[ -1-\frac{1}{\beta}-\frac{\gamma}{\beta} \right] }
	\left( \frac{R}{R_0} \right)^{-\beta/2},
& \quad \beta < 0, \\
	\langle v \rangle &= v_0 \sqrt{\frac{2}{1+\gamma}} 
	\frac
	{\Gamma \left[ 1+\frac{\gamma}{2} \right] }
	{\Gamma \left[ \frac{1}{2}+\frac{\gamma}{2} \right]},
& \quad \beta =0.
\end{xalignat}
The radial velocity dispersion $\sigma_u$ is the root-mean-squared
radial velocity of the stars. We shall also find it convenient to
define a dimensionless radial velocity dispersion $\sigutil$. These
are:
\begin{xalignat}{2}
	\sigma_u^2 &= {{v_\beta^2} \over {1+\gamma+2\beta} } \left( {R_0 \over R} \right) ^{\beta},
	\label{eq:sigmau}
&\quad
	\sigutil^2 &= \left( {R \over R_0} \right) ^{\beta}
 { {\sigma_u^2} \over {v_\beta^2} } 
	= { 1 \over {1+\gamma+2\beta} }
	\label{eq:sigutil}.
\end{xalignat}
If all the stars move on circular orbits, the radial velocity
dispersion vanishes. Such a disk is said to be ``cold'', by analogy
with the motion of molecules in gases. As the eccentricity of the
orbits increases, the stars acquire random motions and thus the
``temperature" of the disk increases.  High values of the anisotropy
parameter $\gamma$ correspond to low velocity dispersions, i.e., cold
disks. This is obvious on examining the form of the distribution
function: $\fsing \varpropto L_z^\gamma$. When $\gamma$ is large, more
of the stars have high angular momentum and lie on near-circular
orbits. The isotropic model is given by $\gamma=0$.  The squared
tangential velocity dispersion $\sigma_v^2$ is the difference between
the second tangential velocity moment $\langle v^2 \rangle$ and the
square of the mean streaming velocity $\langle v \rangle^2$. The
second moment $\langle v^2 \rangle$ is derived from the distribution
function as:
\begin{equation}
	\langle v^2 \rangle
	= {{v_\beta^2 (1+\gamma)} \over {1+\gamma+2\beta} } 
        \left( {R_0 \over R} \right) ^{\beta}
	= (1+\gamma) \sigma_u^2
	.
	\label{eq:meanofvsq}
\end{equation}
As $\gamma \rightarrow \infty$, $\sigma_u^2 \rightarrow 0$ and
$\sqrt{\langle v^2 \rangle} \rightarrow \vcirc$, so the disk is
rotationally supported. These distribution functions have the property
that, at any spot, the ratio of radial velocity dispersion to mean
squared tangential velocity is constant.

It has been conjectured~\cite{OstPeeb:1973} that the global stability
of disks is related to the ratio of the total rotational energy to the
total potential energy. As part of our aim is to examine this claim,
let us derive the global virial quantities for future reference.  The
total kinetic energy $K(R)$ and potential energy $W(R)$ within any
radius $R$ can be computed as
\begin{xalignat}{2}
	K(R) &= \pi\Sigma_0 v_\beta^2 R_0^2
	\frac{2+\gamma}{(1+\gamma+2\beta)(1-2\beta)}
	\left( \frac{R}{R_0} \right)^{1-2\beta},
	\label{eq:kineticT}
&\quad
	W(R) &= -\frac{2\pi\Sigma_0 v_\beta^2 R_0^2}{1-2\beta}
	\left( \frac{R}{R_0} \right)^{1-2\beta}.
	\label{eq:potentialW}
\end{xalignat}
For the power-law disks, the virial theorem takes the form
\be
	2K(R) + \frac{2+\gamma}{1+\gamma+2\beta} W(R) = 0.
	\label{eq:virial}
\end{equation}
Note that the power-law disks do not, in general, satisfy the standard
virial theorem $2K+W=0$. This is because it is not possible to
``enclose'' the system with a sufficiently large container.  No matter
how large the container, if the disk is warm, some stars will always
cross its surface. When the disk is perfectly cold, the stars have no
radial motion and thus do not cross the surface. In this case, as seen
from eq.~\eqref{eq:virial}, the standard virial theorem does hold.
(Similar comments hold good for the isothemal sphere, for example).
The rotational kinetic energy is
\begin{xalignat}{2}
	T(R) &= 
	\frac{2\pi\Sigma_0 v_\beta^2 R_0^2}{\beta(1-2\beta)}
	\frac
	{\Gamma^2\left[ 1+\frac{\gamma}{2} \right]
	 \Gamma^2\left[ 2+\frac{1}{\beta}+\frac{\gamma}{\beta} \right]}
	{\Gamma^2\left[ \frac{1}{2}+\frac{\gamma}{2} \right]
	\Gamma^2\left[ \frac{5}{2}+\frac{1}{\beta}+\frac{\gamma}{\beta} \right]}
	\left( \frac{R}{R_0} \right)^{1-2\beta},\label{eq:rotationT1}
& \quad \beta > 0,\\
	T(R) &= 
	-\frac{2\pi\Sigma_0 v_\beta^2 R_0^2}{\beta(1-2\beta)}
	\frac
	{\Gamma^2\left[ 1+\frac{\gamma}{2} \right]
	 \Gamma^2\left[ -\frac{3}{2}-\frac{1}{\beta}-\frac{\gamma}{\beta} \right]}
	{\Gamma^2\left[ \frac{1}{2}+\frac{\gamma}{2} \right]
	\Gamma^2\left[ -1 -\frac{1}{\beta} - \frac{\gamma}{\beta} \right]}
	\left( \frac{R}{R_0} \right)^{1-2\beta},\label{eq:rotationT2}
& \quad \beta < 0,\\
	T(R) &= 	
	\frac{2\pi\Sigma_0 v_0^2 R_0^2}{1+\gamma}
	\frac
	{\Gamma^2\left[ 1+\frac{\gamma}{2} \right] }
	{\Gamma^2\left[ \frac{1}{2}+\frac{\gamma}{2} \right]}
	\left( \frac{R}{R_0} \right),
& \quad \beta = 0.
	\label{eq:rotationT3}
\end{xalignat}

This concludes our summary of the equilibrium distribution functions
for the power-law disks. In passing, let us emphasise that they are
just the simplest choice of distribution functions, not the only
ones. To illustrate this, Appendix~\ref{sec:SingEcc} gives an example
of a very different set of distribution functions. These perhaps fall
into the class of remarkable curiosities, since the disks are built
from orbits of one shape only.


\subsection{The Cut-Out Distribution Functions}\label{sec:cutouts}

\noindent
So far we have considered only the self-consistent case.  We also plan
to examine disks where parts of the central density are carved
out. This is very much in the spirit of Zang's~\cite*{Zang:1976}
pioneering investigations. The cut-out mass is still present, in the
sense that it contributes to the forces experienced by the remaining
stars, but it is not free to participate in the perturbation. The disk
is thus divided into ``active'' and ``inactive'' components.  Although
motivated partly by mathematical convenience, this is also a
physically reasonable step to take. Stars in galactic disks are
subject not merely to the disk's gravity field, but also to forces
from the halo and bulge. A self-consistent distribution function, such
as $\fsing$, is appropriate only when the disk's self-gravity
overwhelms the gravitational potential of the other components. The
immobile central mass can be interpreted physically as the hot bulge
at the centre of disk galaxies. Another possibility -- suggested to us
by Tremaine (1997, private communication) -- is to interpret the rigid
density as caused by stars on highly elongated radial orbits. They
pass through the centre of the disk, but they spend most of their time
sufficiently far away from the disk so that they do not respond to the
changing potential.  There is one further motivation to consider
cut-out disks as well as self-consistent ones. We wish to compare our
results with those from $N$-body studies -- for example, those of
Sellwood, Earn and collaborators~\cite{SellAna:1986,Earn:1993,ES:1995}.  
Obviously, numerical simulations cannot cope with infinite forces so
in these studies the singularity at the origin is softened. Although
some simulations (e.g., using Bessel function expansions) can track
the behaviour of stars to infinite distances, disks with infinite mass
are still problematic. For this reason, the disk is truncated at some
finite radius. For comparison with these studies, an outer cut-out is
needed, as well as an inner one.

To ensure that we do not run afoul of Jeans'~\shortcite{Jeans:1919}
theorem, the carving-out is performed by multiplying our self-consistent
distribution function $\fsing$~\eqref{eq:fsing} by a cut-out function
$H(L_z)$:
\begin{equation}
	\fcutout (E,L_z) = H(L_z) \fsing(E,L_z).
	\label{eq:fcutout}
\end{equation}
We carry out our analysis for three cut-out functions $H(L_z)$: the
self-consistent (scale-free) disk, the inner cut-out disk and the
doubly cut-out disk. These are given respectively by
\begin{xalignat}{3}
	H(L_z) & = 1,
&\quad
	H(L_z) = { {\LNi} \over {\LNi + \left( v_\beta R_0 \right)^{N_\beta} } },
&\qquad
	H(L_z) = { {\LNi \Lc^{M_\beta}} \over 
{ \left[\LNi + \left( v_\beta R_0 \right)^{N_\beta}\right] 
\left[ \LNo + \Lc^{M_\beta} \right] } } 
	\label{eq:innercutoutfunction}	 
	\label{eq:doublycutoutfunction}.
\end{xalignat}
In Section 3, we will frequently use an equivalent form of the cut-out
function expressed in dimensionless variables defined by $\Htil
(\Ltil) \equiv H(L)$.  In all these formulae,
\begin{xalignat}{2}
 N_\beta & = { {2 + \beta} \over {2 - \beta } } N,
 & \qquad M_\beta &  = { {2 + \beta} \over {2 - \beta } } M  ,
	\label{eq:NbetaMbeta}
\end{xalignat}
where $N$ and $M$ are the inner and outer cut-out indices,
respectively and must be positive integers.  The choice of the inner
cut-out reduces to Zang's~\cite*[eq. Z2.57]{Zang:1976} when $\beta
\rightarrow 0$. Our generalisation seems to come ``out of thin air'', 
but we shall see in Section 3.5 and Appendix~\ref{sec:FlmIntegration}
that our choice enables the analytic, rather than numerical,
evaluation of a contour integral.  ~\longcite{Earn:1993} uses a doubly
cut-out function of the same form as ~\eqref{eq:doublycutoutfunction}
to carry out his numerical simulations (although, not having to
perform any contour integrals, Earn is free to choose non-integral $N$
and $M$).

What effect does this modification of the distribution function have
on the surface density? When the disk is cold, the cut-out function
depends only on radius and the active surface density may be
calculated exactly. For a doubly cut-out disk, it is
\begin{equation}
	\frac{ \Sigma_{\text{active}}^{\text{cold}}}{\Sigeq} = 
	\frac
	{ R^{ \frac{2+\beta}{2} N} }
	{ \left[ R^{ \frac{2+\beta}{2} N} +  R_0^{ \frac{2+\beta}{2} N}  \right] }
	\frac
	{ (R_0 \Rtilc )  ^{ \frac{2+\beta}{2} M} }
	{ \left[ R ^{ \frac{2+\beta}{2} M} 
	+  (R_0 \Rtilc )  ^{ \frac{2+\beta}{2} M} \right] }
	\label{eq:colddoublycutout},
\end{equation}
where the truncation radius $\Rtilc$ is given by $ \Rtilc =
\Ltilc^{2 / (2 - \beta )} $.  The proportion of the
equilibrium disk which remains active rises from zero at the centre of
the disk to unity at larger radii. At $R = R_0$, the inner cut-out
removes exactly half the equilbrium density. The value of $N$ controls
the steepness of the rise: the cut-out is gentler for smaller values
of $N$.  Conversely, the outer cut-out, parametrised by $M$, removes
matter from the outer regions of the disk. At $R = \Rtilc R_0$, it
removes half the equilibrium density, with $M$ controlling the
sharpness of the cut-out.

\begin{figure}
	\begin{center}
	\epsfig{file=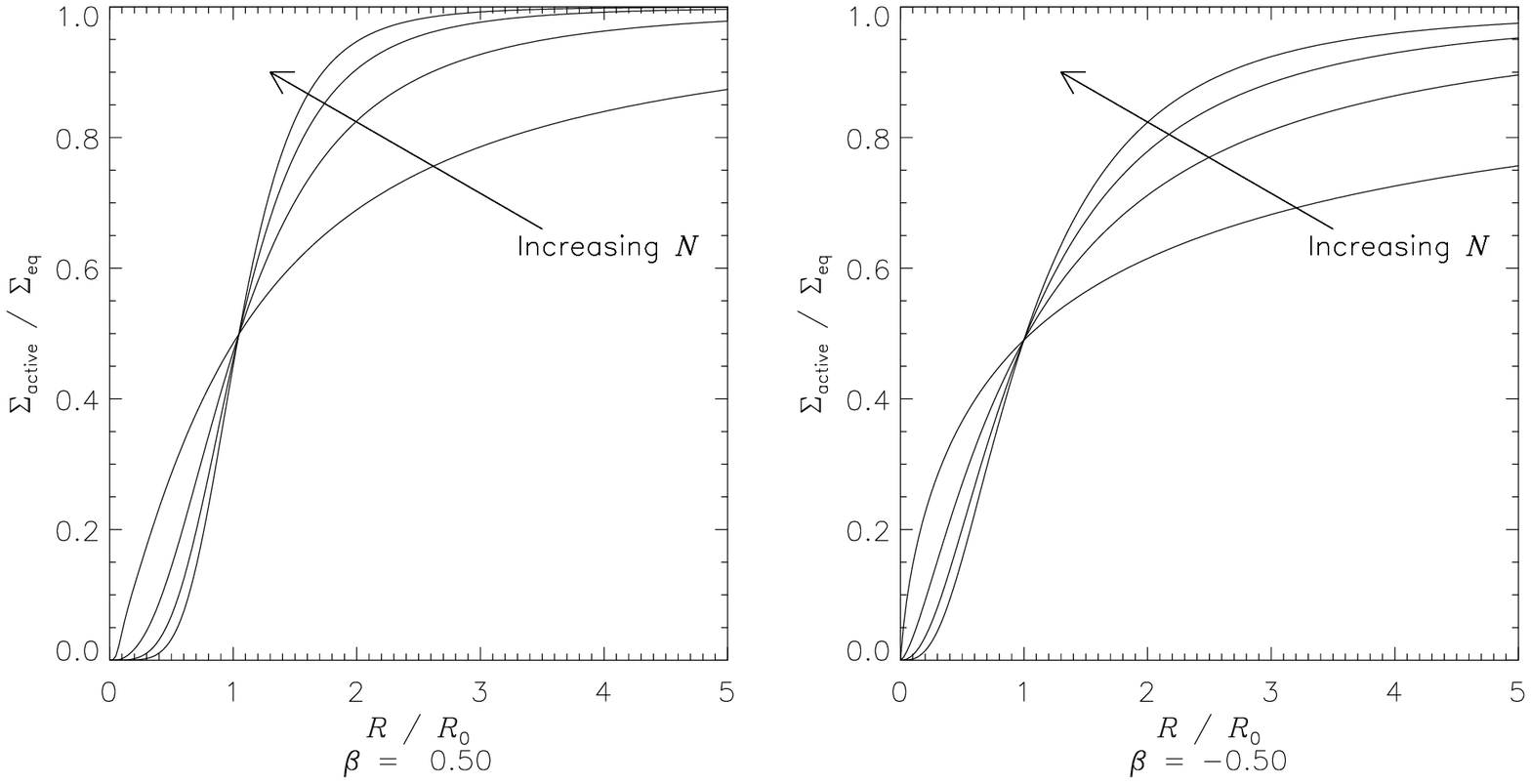,width=0.8\textwidth,height=0.3\textwidth}
	\caption[Fractional active surface density of a hot disk with
	an inner cut-out function] {The active surface density of a
	hot disk with an inner cut-out function, shown as a fraction
	of the equilibrium density. The cut-out becomes more abrupt
	with increasing $N$. [The solid lines are the density for
	$N=1$, 2, 3, 4. For $\beta=+0.5$, $\sigutil=0.199$; for
	$\beta=-0.5$, $\sigutil=0.740$. These are the temperatures at
	which the self-consistent disk is locally stable to
	axisymmetric disturbances.]
\label{fig:InnerFracDensity}
}
		\epsfig{file=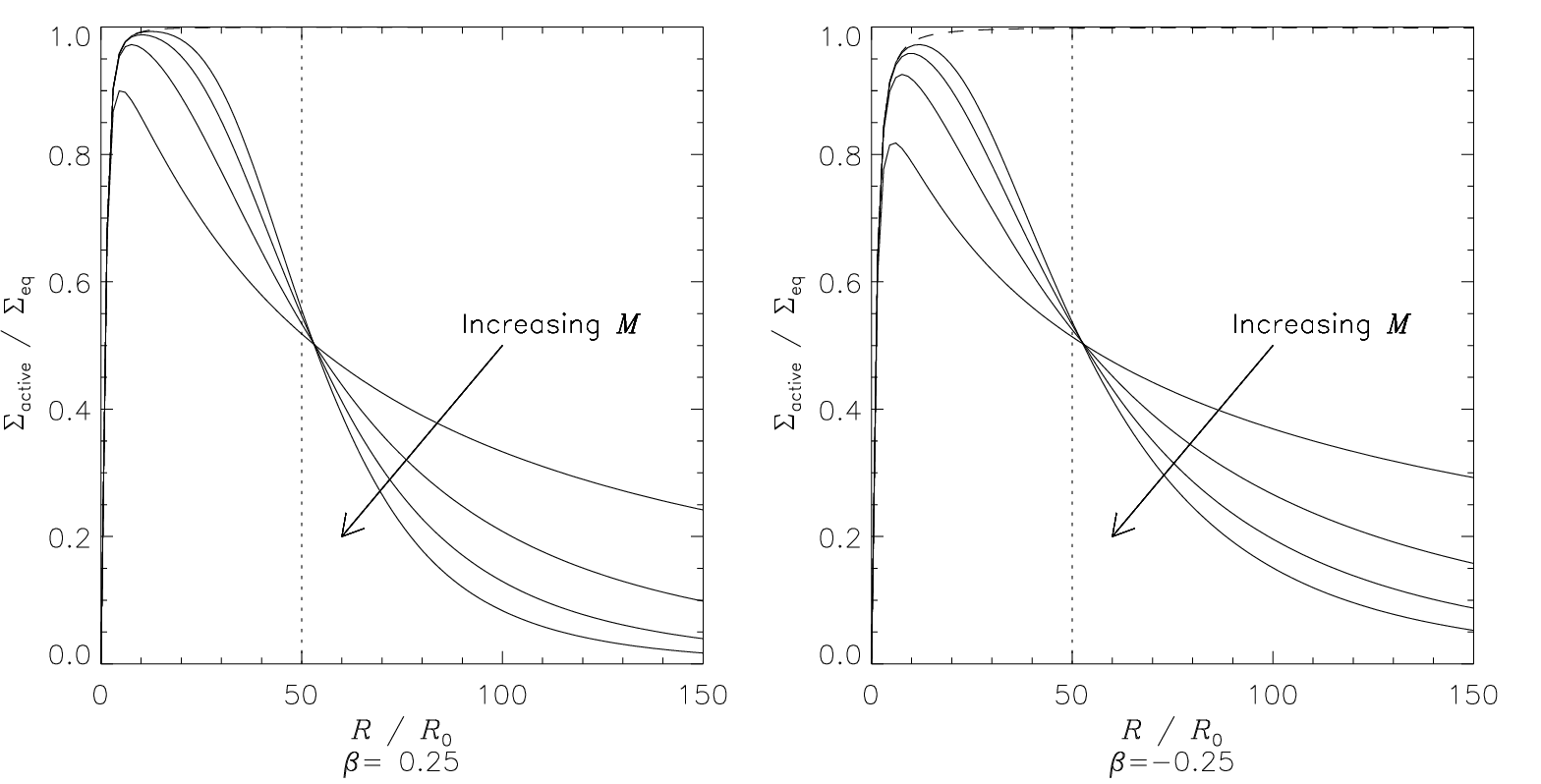,width=0.8\textwidth,height=0.3\textwidth}
		\caption[Fractional active surface density of a hot
		disk with a doubly cut-out function] {The active
		surface density of hot disks with a doubly cut-out
		function, shown as a fraction of the equilibrium
		density. The arrows indicate the direction of
		increasing $M$, or increasing severity of the
		truncation.  The dashed line just visible at the top
		of the plots is the corresponding inner cut-out disk.
		[In each case, $N=2$, $\Rtilc=50$. The solid lines are
		the density for $M=1$, 2, 3, 4. The dotted line marks
		the position of $\Rtilc$. For $\beta=+0.25$,
		$\sigutil=0.283$; for $\beta=-0.25$,$\sigutil=0.509$.]
\label{fig:OuterFracDensity}
}
	\end{center}
\end{figure}

The active surface density of a hot disk can be calculated by
numerical integration. In fact, heating the disk makes little
difference to the active density.  Figs.~\ref{fig:InnerFracDensity}
and~\ref{fig:OuterFracDensity} show the active surface density of
inner and doubly cut-out hot disks. In each case, the active surface
density is expressed as a fraction of the equilibrium surface density.
The form of these curves is close to that of $\Htil(\Ltil)$. Thus the
value of $\Htil(\Ltil)$ is a good approximation to the proportion of
density which is active at $R=R_0 \Ltil$.  Our motivation for
introducing the outer cut-off is to enable our results to be directly
compared against $N$-body work. In practice, the outer cut-out does
not usually have a significant effect on the stability properties of
the disk (unless it is so sharp as to provoke edge modes).

\section{The Fredholm Integral Equation}\label{sec:IntEqnChap}
\noindent
This section derives the Fredholm integral equation for the linear
normal modes of the power-law disks. The first step is to decompose
the imposed density into logarithmic spirals, as in Section 3.1.
Linearising the collisionless Boltzmann equation enables us to
calculate the density response.  The condition for a self-consistent
normal mode is that the imposed density equals the response
density. This yields the Fredholm integral equation written down in
Section 3.2.  The kernel of the integral equation is the transfer
function, describing how an excitation at any wavenumber feeds into
response at all other wavenumbers. The general formulae for the
transfer function is derived in Section 3.3. It depends on the
properties of the equilibrium disk through the angular momentum
function, as discussed in Section 3.4.

%
\subsection{The Logarithmic Spirals\label{sec:logspir}}
\noindent
The logarithmic spirals were made famous by Snow~\cite*{Snow:1952} and
Kalnajs~\cite*{Kalnajs:1965,Kalnajs:1971}. They have surface density
\begin{equation}
	\Sigimp^{\alpha m}
	 = 
	\Sigmap  e ^{st} e ^{ im\left( \theta - \Omegap t\right) }
	 \left( { {R_0} \over {R} } \right) ^ {3/2 - i\alpha}
	 = 
	\Sigmap  e ^{ i\left( m\theta - \omega t\right) }
	 \left( { {R_0} \over {R} } \right) ^ {3/2 - i\alpha},
\label{eq:dencomponent}
\end{equation}
where $\Sigmap$ is a constant amplitude. This $m$-lobed pattern
rotates with a constant pattern speed $\Omegap$ and grows or decays
with a growth rate $s$.  The logarithmic wavenumber (hereafter
abbreviated to just wavenumber) $\alpha$ controls how tightly the
spiral is wrapped. The pattern is more tightly wound for larger values
of $\alpha$. It is often convenient to collect the growth rate and
pattern speed into the complex frequency $\omega = m\Omegap + is$.  By
adding logarithmic spirals with different $\alpha$, we can build up general
patterns with azimuthal wavenumber $m$:
\be
	\Sigimp \left( R, \theta \right) = \int_{-\infty}^{+\infty} d\alpha \Aimp \left( \alpha \right) \Sigimp^{\alpha m} .
	\label{eq:builddensity}
\end{equation}
Any imposed density perturbation $\Sigimp$ is expanded in a Fourier
series of azimuthal harmonics characterised by $m$.  In the linear
r\'egime, the response of the disk to the change in potential has the
same order $m$ of rotational symmetry as the imposed density
perturbation. This means that our investigations can be confined to a
single value of $m$ at a time.  The expansion in logarithmic spirals is
equivalent to taking the Fourier transform in the variable $\xtil=\ln
(R/R_0).$ This is evident on defining
\be
\Sigma_m  =  \Sigmap e ^{ i\left( m \theta - \omega t\right) } e^{-3\xtil/2}	\label{eq:Sigmam}
\end{equation}
so that we obtain the conventional Fourier transform pair:
\begin{xalignat}{2}
	{ {\Sigimp  } \over { \Sigma_m} } & =
 	\int_{-\infty}^{+\infty} d\alpha \, \Aimp \left( \alpha \right) e^{ i\alpha \xtil},
\label{eq:transformpairSigma}
& \quad
 	\Aimp \left( \alpha \right) & = { 1\over {2 \pi } }
	\int_{-\infty}^{+\infty}d\xtil \, { { \Sigimp} \over { \Sigma_m } }e^{  - i\alpha \xtil  } 	.
\label{eq:transformpairA}
\end{xalignat}
Here, $\Aimp$ is the Fourier transform of the imposed density
perturbation. For the transform to exist, $\Sigimp \left( R, \theta
\right) R^{3/2}$ must tend to zero as $R \rightarrow 0$ and $R
\rightarrow \infty$. This means that $\Sigimp \left( R, \theta 
\right)$ can diverge at the centre no faster than $R^{-3/2}$ and 
must fall off at large radii more quickly than $R^{-3/2}$.
Kalnajs~\cite*{Kalnajs:1965,Kalnajs:1971} derived the potential
$\psimp^{\alpha m} $ corresponding to a single logarithmic spiral
component $\Sigimp^{\alpha m}$ as
\be
	  \psimp^{\alpha m}   
	=  2 \pi G \Sigmap K(\alpha,m) R_0
	e^{ i \left( m \theta - \omega t \right) }
	\left( {R \over {R_0}} \right) ^{ i  \alpha - \half }
	\label{eq:psialpham},
\end{equation}
where $K(\alpha,m)$ is the Kalnajs gravity function
\be
 	K(\alpha,m) = \half { {
	\Gamma \left[ \half \left( \half + m + i \alpha \right)  \right]
	\Gamma \left[ \half \left( \half + m - i \alpha \right)  \right]
	 } \over {
	\Gamma \left[ \half \left( {3 \over 2}+ m + i \alpha \right)  \right]
	\Gamma \left[ \half \left( {3 \over 2}+ m - i \alpha \right)  \right]
	} }.
	\label{eq:KGRAV}
\end{equation}
This function is real and positive for real $\alpha$. It has the
symmetry $K(\alpha,m) = K(-\alpha,m)$.

From the imposed potential $\psimp^{\alpha m}$, we obtain the change
in the distribution function $\fimp^{\alpha m}$.  The linearised
collisionless Boltzmann equation~\cite[chap. 5]{BnT} relates the
change in the distribution function to the changes in energy and
angular momentum experienced by individual stars as a result of the
density perturbation:
\begin{equation}
	{ {d \fimp} \over {d t} } 
= 
	 -\left( u { {\partial \psimp} \over {\partial R} } 
	+ {v \over R} { {\partial \psimp} \over {\partial \theta} } \right)
	{ {\partial f} \over {\partial E} }
	- { {\partial \psimp} \over {\partial \theta} }
	 { {\partial f} \over {\partial L_z} }
=
	- { {\partial f} \over {\partial E} } {{dE} \over {dt} }
	- { {\partial f} \over {\partial L_z} } {{dL_z} \over {dt}} 
	\label{eq:ddt_pertbn}.
\end{equation}
In physical terms, eq.~\eqref{eq:ddt_pertbn} simply states that the
rate at which stars move to perturbed orbits is equal to the rate at
which stars leave equilibrium orbits. Integrating~\eqref{eq:ddt_pertbn} 
over the entire time of the perturbation, we obtain
\be
	\fimp^{\alpha m}  (t) = - { {\partial f} \over {\partial E} } \Delta E
	- { {\partial f} \over {\partial L_z} } \Delta L_z 
	\label{eq:fimp},
\end{equation}
where $\Delta E$ and $\Delta L_z$ are the changes in the star's energy and
angular momentum due to the perturbation. For a single logarithmic
spiral component, these simplify to
\begin{xalignat}{2}
	\Delta E 
	= \int_{-\infty}^{t} 
	\left(
	u^{\prime} 
	{{\partial \psimp^{\alpha m} } \over {\partial R^{\prime}} } 
	+ {v^{\prime} \over R^{\prime}}
	{{\partial \psimp^{\alpha m} } \over {\partial \theta^{\prime}} } 
	\right)
	dt^{\prime}\label{eq:DeltaE}
	& =  \psimp^{\alpha m}  (t) - \int_{-\infty}^{t}{ {\partial \psimp^{\alpha m}  } \over {\partial t^\prime } }dt^{\prime}.
&\qquad
	\Delta L_z 
	& = \int_{-\infty}^{t} {{\partial \psimp^{\alpha m} } \over {\partial \theta^{\prime}} } dt^{\prime}
	\label{eq:DeltaL_z}.
\end{xalignat}
Physically, $\Delta E$ is the difference between the potential here
and now, and an averaged potential sampled by the orbit over its
history.  Note that this derivation assumes that the perturbation
vanished in the distant past.


\subsection{The Integral Equation\label{sec:IntEqn}}
To find the change in surface density $\Sigres^{\alpha m}$ caused by a
single logarithmic spiral component $\Sigimp^{\alpha m}$, an integration of
$\fimp^{\alpha m}$ must be performed over all velocities $u$ and $v$
\be
	\Sigres^{\alpha m} = \iint \fimp^{\alpha m} \, du \, dv.
\end{equation}
To find the total change in density caused by the whole disturbance
$\Sigimp = \int d\alpha \, \Aimp \left( \alpha \right) \Sigimp^{\alpha
m}$, we integrate over all the logarithmic spiral components
\be
	\Sigres = \int_{-\infty}^{+\infty} d\alpha \, \Aimp \left( \alpha \right) \Sigres^{\alpha m}.
	\label{eq:Sigmares1}
\end{equation}
It is possible to equate $\Sigres$ and $\Sigimp$ and seek to solve the
resulting equation for the self-consistent solutions.  A more
profitable approach is to equate the density transforms. We define the
response density transform $\Ares \left( \alpha \right)$ analogously 
to~\eqref{eq:transformpairSigma} so that
\begin{xalignat}{2}
	{ {\Sigres } \over { \Sigma_m} }&=
 	\int_{-\infty}^{+\infty} d\alpha \, \Ares \left( \alpha \right) e^{ i\alpha  \xtil } ,
	\label{eq:Sigmares}
&\quad
	\Ares \left( \alpha \right) &= { 1\over {2 \pi } }
	\int_{-\infty}^{+\infty}d\xtil \, 
	{ { \Sigres  } \over { \Sigma_m } }e^{  - i\alpha  \xtil }.
	\label{eq:Ares}
\end{xalignat}
Substituting for $\Sigres$ from~\eqref{eq:Sigmares1}, the response
density transform becomes
\be
	\Ares \left( \alpha \right) = { 1\over {2 \pi } }
	\int_{-\infty}^{+\infty}d\xtil \,
	{{ e^{  - i\alpha \xtil } } \over {	\Sigma_m } }
	  \int_{-\infty}^{+\infty} d\alpha^\prime \Aimp \left( \alpha^\prime \right) \Sigres^{\alpha^\prime m}.
\end{equation}
Exchanging the order of integration, we obtain (cf. Z3.42):
\begin{xalignat}{2}
	\Ares \left( \alpha \right) 
	&=
	\int_{-\infty}^{+\infty} d\alpha^\prime
	\Aimp \left( \alpha^\prime \right) 
	\Sm \left( \alpha, \alpha^\prime \right) ,
\label{eq:inteqn}
&\quad 
\text{where}	\hspace{1cm}
 	\Sm \left( \alpha, \alpha^\prime \right) = { 1\over {2 \pi } }
	\int_{-\infty}^{+\infty} d\xtil \, 
	{ {\Sigres^{\alpha^\prime m}} \over { \Sigma_m } }
	 e^{  - i\alpha \xtil }.
\label{eq:Sm}
\end{xalignat}
Self-consistency requires the response density to equal the imposed
density, or equivalently $\Aimp \left( \alpha \right) = \Ares \left(
\alpha \right)$. Equation~\eqref{eq:Sm} therefore becomes an 
integral equation for the density transform of the normal modes of
oscillation. Of course, similar integral equations have been derived
by others before. For example, Kalnajs~\cite*{Kalnajs:1971} derived a
completely general integral equation in action-angle coordinates,
whereas Palmer \& Papaloizou~\cite*{PalPap:1990} restricted their
attention to disks built from epicyclic orbits. This form of the
integral equation was first derived by Zang~\cite*{Zang:1976}.  The
kernel of the integral equation $\Sm \left( \alpha, \alpha^\prime
\right)$ is called the {\it transfer function} and describes how much of
the disks response to the disturbance with wavenumber $\alpha^\prime$
occurs at the wavenumber $\alpha$. It is to the transfer function that
our attention now turns.


\subsection{The Transfer Function	\label{sec:TransferFunct}}

\begin{figure}
	\begin{center}
		\epsfig{file=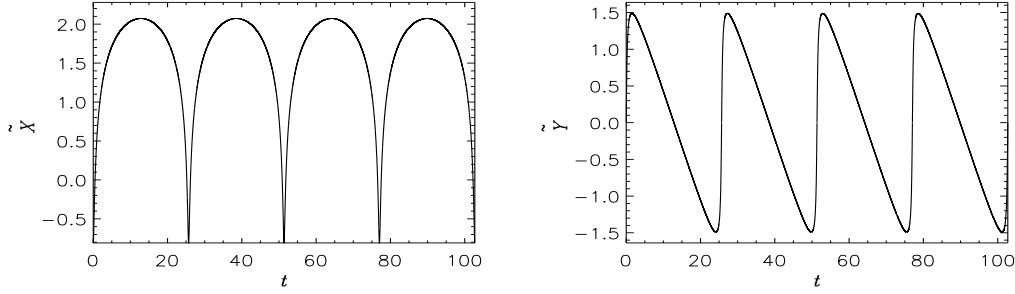,width=0.8\textwidth,height=0.25\textwidth}
		\caption[Excursions of a stellar orbit] {Excursions of
		a stellar orbit as viewed in the $\Xtil$ and $\Ytil$
		coordinates. The left panel shows the scaled
		logarithmic radius $\Xtil$ and the right panel the
		deviation from mean angular motion $\Ytil$ plotted
		against time. [Numerical details: $U = 1.5$, $R_H =
		1$, $\beta=0.25$, $R_0 = v_\beta = 1$.]\label{fig:XY}}
		\end{center}
\end{figure}

\noindent
This section contains an elaborate calculation of the transfer
function of the power-law disks. To aid the reader, a reference table
of some of the quantities are collected in Appendix~\ref{sec:RefTabs}.
The first step is to evaluate the change in distribution function
$\fimp^{\alpha m}$ given in eq.~\eqref{eq:fimp}. From eqs.
~\eqref{eq:psialpham} and ~\eqref{eq:DeltaL_z}, the changes in energy
and angular momentum caused by a single logarithmic spiral component
are
\begin{xalignat}{2}
	\Delta E & = \psimp^{\alpha m} (t) + i \omega \int_{-\infty}^{t}\psimp^{\alpha m}\left( t^\prime \right) dt^{\prime}	 \label{eq:DeltaEpsi},
 &
\quad
	\Delta L_z & = im \int_{-\infty}^{t} \psimp^{\alpha m} \left( t^\prime \right) dt^{\prime}	 \label{eq:DeltaLpsi}.
\end{xalignat}
The integration is simplified by shifting to the frame which rotates
at the star's average velocity $\Omega$. In this frame, the star's
orbit closes and the integrands in eq.~\eqref{eq:DeltaLpsi} become
periodic with the radial period $T$.  Following Zang~\cite*{Zang:1976}, 
let us define two dimensionless coordinates describing the periodic 
excursions of the stellar orbits as:
\begin{xalignat}{2}
	\Xtil &= \ln {R \over {R_H}} = \ln {\Rtil},
& \quad
	\Ytil &= \theta - \Omega t = \theta - \tilde{\Omega} \tilde{t}.
\end{xalignat}
Fig.~\ref{fig:XY} shows how $\Xtil$ and $\Ytil$ vary over four radial periods
for a somewhat eccentric orbit.
The potential $\psimp$~\eqref{eq:psialpham} experienced by the star at 
time $t^\prime$ when its position is $( R^\prime, \theta^\prime )$, or
equivalently $\Xtil^\prime = \ln \Rtil^\prime $, $\Ytil^\prime = 
\theta^\prime-\Omega t^\prime$, is:
\begin{equation}
	\psimp^{\alpha m} ( t^\prime ) = 
	2 \pi G \Sigmap K(\alpha,m) R_0 \Rtil_H ^{ i  \alpha - \half }
	\exp
	\left\{ 
	i( m \Omega  -\omega ) t^\prime  
	+ im\Ytil^\prime + ( i  \alpha - \fr12 ) \Xtil^\prime 
	\right\}.
	\label{eq:001}
\end{equation}
Now $\Xtil$ and $\Ytil$ both have period $T$, so the term 
$ \exp \left\{ im\Ytil^\prime + ( i \alpha - \fr12 ) \Xtil^\prime \right\}$
is similarly periodic as $t^\prime$ varies.  It can therefore be
expanded in a Fourier series (a method previously used by
Kalnajs~\cite*{Kalnajs:1972} and Zang~\cite*{Zang:1976}):
\begin{equation}
	\exp
	\left\{
	im\Ytil^\prime + ( i  \alpha - \fr12 ) \Xtil^\prime 
	\right\}
	=
	\sum_{l=-\infty}^{+\infty} \Qlm(\alpha) 
	\exp\left\{ {{2i\pi l} \over T}t^\prime \right\},
	\label{eq:010}
\end{equation}
where the Fourier coefficient $\Qlm (\alpha)$ is given by
\begin{equation}
	\Qlm(\alpha) ={1 \over T} \int_0^T
	\exp \left\{
	im\Ytil^\prime + ( i  \alpha - \fr12 ) \Xtil^\prime 
	-{{2i\pi l} \over T}t^\prime 
	\right\}
	dt^\prime.
\end{equation}
Changing variables to the orbital phase $\chi=\kappa t$, the
Fourier coefficient becomes (Z3.27)
\be
 	\Qlm(\alpha) ={1 \over {2\pi}} 
	\int_0^{2\pi}
	\exp \left\{ im \Ytil^\prime
	+ ( i  \alpha - \fr12 ) \Xtil^\prime -   il \chi^\prime \right\}
	d\chi^\prime
	\label{eq:Qlm}.
\end{equation}
Substituting this into our expression~\eqref{eq:001}, we obtain for
the perturbation potential sampled by the star
\be
	\psimp^{\alpha m} ( t^\prime ) = 
	2 \pi G \Sigmap K(\alpha,m) R_0 \Rtil_H ^{ i  \alpha - \half }
	e^{ i ( m \Omega  -\omega ) t^\prime } 
	\sum_{l=-\infty}^{+\infty} \Qlm(\alpha) e^{i l \kappa t^\prime }.
\end{equation}
Using~\eqref{eq:DeltaEpsi}, the changes in the energy and the angular
momentum of the star caused by a single logarithmic spiral is
\begin{equation}
	\Delta E  = 2 \pi G \Sigmap K(\alpha,m) R_0 \Rtil_H ^{ i  \alpha - \half } 
	e^ {i ( m \Omega  -\omega ) t}
	\sum_{l=-\infty}^{+\infty} \Qlm(\alpha) e^{i l \chi }
	{ { l \kappa +m \Omega } \over { l \kappa +m \Omega -\omega } }
	\label{eq:DeltaEQlm},
\end{equation}
\begin{equation}
	\Delta L_z = 2 \pi m G \Sigmap K(\alpha,m) R_0 \Rtil_H ^{ i  \alpha - \half }
	e^{ i ( m \Omega  -\omega ) t }
	\sum_{l=-\infty}^{+\infty}
	 { {\Qlm(\alpha) e^{ i l \chi } } \over {l \kappa +m \Omega -\omega } }
	\label{eq:DeltaLQlm}.
\end{equation}
These correspond to (Z3.29) and (Z3.30), although Zang uses a slightly
different form of the Fourier coefficient. Substituting these
expressions into~\eqref{eq:fimp} gives the final expression for the
change in the distribution function:
\begin{equation}
	\fimp^{\alpha m} (t) = -2 \pi G \Sigmap K(\alpha,m) R_0 \Rtil_H ^{ i  \alpha - \half }
	e^{ i ( m \Omega  -\omega ) t }
	\sum_{l=-\infty}^{+\infty}
	{ {\Qlm(\alpha) e^{i l \chi }} \over { l \kappa +m \Omega -\omega } }
\left\{
	( l \kappa +m \Omega  ) { {\partial f} \over {\partial E} }
	+  m { {\partial f} \over {\partial L_z} } 
\right\}.
	\label{eq:genfimp}
\end{equation}
Thus far, our results are independent of the form of the equilibrium
disk. Let us now specialise to the power-law disks by substituting for the
derivatives of the distribution function~\eqref{eq:fcutout}.  Bearing
in mind that the energy $E$ has the opposite sign to $\beta$, we can
combine the derivatives obtained for positive and negative $\beta$.
\begin{xalignat}{2} { {\partial f}
\over {\partial E} }& = - \tilde{C} \left| \betagammabig \right|
H(L_z) L_z^{\gamma} |E|^{\betagammasmall -1} ,
\label{eq:dfde}
&\quad
{ {\partial f} \over {\partial L_z} } & = 
	 \tilde{C} L_z^{\gamma} |E|^{\betagammasmall}
	\left( \gamma { {H(L_z)} \over L_z} + { {dH} \over {dL_z}} \right).
\label{eq:dfdl}
\end{xalignat}
We also replace the dimensional quantities with their dimensionless
analogues (eqs.~\eqref{eq:ELnondim},~\eqref{eq:kaptil},~\eqref{eq:Omtil})
\begin{xalignat}{3}
	L_z & = v_\beta R_0 \Ltil,
& \quad
	 \frac{dH}{dL_z} & = \frac{1}{ v_\beta R_0} \frac{d\Htil}{d\Ltil} , 
& \quad
	E & = \Etil v_\beta^2 \Ltil^{ {{2\beta} \over {\beta-2}}},
\\
	\kappa & = \kaptil {{v_\beta} \over {R_0}} \Ltil^{\mtwopm},
& \quad
	\Omega & = \Omtil {{v_\beta} \over {R_0}} \Ltil^{\mtwopm},
& \quad
	\omega &= \omtil {{v_\beta} \over {R_0}},
\end{xalignat}
Note that $\omtil$ is defined slightly differently from $\kaptil$ and
$\Omtil$. With the present definitions, all three dimensionless
frequencies are independent of $\Ltil$ (it is clear from
eqs.~\eqref{eq:kaptil}, ~\eqref{eq:Omtil} and~\eqref{eq:genintI} that
$\kaptil$ and $\Omtil$ depend only on $\Util$). With these
substitutions, the change in the distribution function now becomes
\begin{equation}
\begin{split}
	\fimp^{\alpha m} (t) &  = 2 \pi G \Sigmap K(\alpha,m) \tilde{C} 
	R_0^{\gamma+1}
	v_\beta^{{2 \over \beta} (1+\gamma) -2 }
	e^{ i ( m \Omega  -\omega ) t } 
	e^{ i  \alpha {2 \over {2-\beta}} \ln\Ltil }
	\Ltil^{ {2\beta -3} \over {2-\beta }}
	| \Etil | ^{\betagammasmall}
\\ & \times 
	\sum_{l=-\infty}^{\infty}
	{ {\Qlm(\alpha) e^{i l \chi }} 
	\over 
	{ \lkapmom - \omtil \Ltil^{\twopm} } 	} 
\left[
	\left\{
	( \lkapmom  )
 \left| \betagammabig \right|  { 1 \over {| \Etil| } }
	- \gamma m
	\right\}
	 \Htil ( \Ltil)
	- m \Ltil \frac{d\Htil}{d\Ltil}
\right].
\label{eq:fimpalpham}
\end{split}
\end{equation}
This is the change in the distribution function brought about by a
single logarithmic spiral component.

Let us now proceed to find the transfer function $\Sm$ from~\eqref{eq:Sm}. 
Substituting for the response surface density in terms of the response
distribution function, we have:
\be 
	\Sm ( \alpha, \alpha^\prime ) = { 1\over {2 \pi } }
	\int_{-\infty}^{+\infty} d\xtil \,\frac{ e^{  - i\alpha \xtil }}
	 { \Sigma_m }  \iint \fimp^{\alpha^\prime m} \, du \, dv.
\end{equation}
This triple integral is transformed to one over the eccentric
velocity $\Util$, the orbital phase $\chi$ and the dimensionless
angular momentum $\Ltil$. A careful calculation of the 
Jacobian~\cite{Read:1997} reveals:
\be
	\frac
	{\partial ( \xtil, u, v ) }
	{\partial ( \chi,\Util,\Ltil ) }
	=
	\frac{v_\beta^2 }{2\pi}
	 \frac{ \auxint_0(\Util)  \Util}{\Rtil}
	 e^{-\xtil} \Ltil^{\beta \over {\beta-2} }.
\end{equation}
The transfer function then becomes
\begin{equation} 
	\Sm ( \alpha, \alpha^\prime ) = 
	{ {v_\beta^2}\over {2 \pi } } 
	\iiint d\Util \, d\chi \, d\Ltil
	{ { e^{  - (i\alpha+1) \xtil }} \over { \Sigma_m } } 	
	{ {  \auxint_0(\Util)} \over {2\pi} } \frac{\Util}{\Rtil}
	\Ltil^{\beta \over {\beta-2} }\fimp^{\alpha^\prime m} (\Ltil,\Util,\chi).
\end{equation}
The angular momentum is integrated from zero to infinity, and the
orbital phase from 0 to $2\pi$. The eccentric velocity is integrated
from zero to infinity for negative $\beta$; for positive $\beta$ the
upper limit is $( 2/\beta - 1)^{1/2}$. For brevity, these limits of
the integration are not shown explicitly.

We shall work in terms of $\Xtil = \ln \Rtil = \ln ( R / R_H )= \xtil
- {2 \over {2-\beta}} \ln \Ltil$. Then $\Rtil = e^{\Xtil}$, and on
substituting for $\Sigma_m $ from~\eqref{eq:Sigmam} we obtain
\begin{equation}
	\Sm ( \alpha, \alpha^\prime ) = 
	{ { v_\beta^2 } \over {2 \pi \Sigmap } } { 1 \over {2\pi} }
	\iiint d\Util \, d\chi \, d\Ltil
	e^{-i(m\theta-\omega t) } 
	e^{ - ( i\alpha+\half ) \Xtil }
	e^{ \frac{1-\beta - 2i \alpha}{2-\beta}  \ln \Ltil }
	\auxint_0(\Util) \Util
	\fimp^{\alpha^\prime m} (\Ltil,\Util,\chi).
	\label{eq:047}
\end{equation}
Note that $t$, $\Xtil$ and $\theta$ describe where the star is in its
orbit. They therefore depend on the orbital phase $\chi$, as well as on
$\Util$ which describes the shape of the orbit.  Substituting for
$\fimp^{\alpha^\prime m}$ from~\eqref{eq:fimpalpham}:
\begin{equation}
\begin{split}
	\Sm ( \alpha, \alpha^\prime ) 
& = 
	 R_0^{\gamma+1}
	v_\beta^{{2 \over \beta} (1+\gamma) }
	 \frac{G  K(\alpha^\prime,m) \tilde{C} }{2\pi}
	\iiint d\Util \, d\chi \, { {d\Ltil} \over {\Ltil} }
	e^{ - im ( \theta - \Omega t ) }
	e^{ - ( i\alpha+\half) \Xtil }
	e^{ {- {2i} \over {2-\beta}} ( \alpha - \alpha^\prime ) \ln \Ltil }
	\Util \auxint_0(\Util) 
	| \Etil | ^{\betagammabig}
\\ & \times 
	\sum_{l=-\infty}^{+\infty}
	{ {\Qlm ( \alpha^\prime ) e^{i l \chi }} 
	\over { \lkapmom -\omtil \Ltil^{\twopm} } } 
\left[
	\left\{
	( \lkapmom )
 \left| \betagammabig \right|  { 1 \over {| \Etil | } }
	- \gamma m
	\right\}
	 \Htil ( \Ltil)
	- m \Ltil \frac{d\Htil}{d\Ltil}
\right] .
\end{split}
\end{equation}
Substituting $\Etil = ( \beta\Util^2 + \beta  - 2 ) / (2\beta )$
and $\theta - \Omega t = \Ytil(\chi)$,
\begin{equation}
\begin{split}
	\Sm ( \alpha, \alpha^\prime ) 
& = 
	R_0^{\gamma+1}
	v_\beta^{{2 \over \beta} (1+\gamma) }
	2 \pi G K(\alpha^\prime,m) \tilde{C} 
	\int d\Util 
	\auxint_0(\Util) \Util
	\modenergy ^{\betagammabig}
\\ & \times
	\sum_{l=-\infty}^{+\infty} \Qlm ( \alpha^\prime )
	{ 1 \over {2\pi} }
	\int d\chi
	e^{ i l \chi -im \Ytil - ( i\alpha+\fr12) \Xtil } 
\\ & \times
	{ 1 \over {2\pi} } \int { {d\Ltil}  \over \Ltil}
	\frac
	{e^{{-2i\over {2-\beta}} ( \alpha - \alpha^\prime ) \ln \Ltil }}
	{ \lkapmom -\omtil \Ltil^{\twopm}} 
\left[
	\left\{
	( \lkapmom )
 	\modterm - \gamma m
	\right\}
	 \Htil ( \Ltil)
	- m \Ltil \frac{d\Htil}{d\Ltil}
\right].
\end{split}
\end{equation}
From the definition of the Fourier coefficient~\eqref{eq:Qlm}, the
integral over orbital phase $\chi$ is just the complex conjugate of
$\Qlm (\alpha)$.  Then, the transfer function becomes:
\begin{equation}
	\Sm ( \alpha, \alpha^\prime )  = 
	R_0^{\gamma+1}
	v_\beta^{{2 \over \beta} (1+\gamma) }
	2 \pi G K(\alpha^\prime,m) \tilde{C} 
	\int d\Util 
	\auxint_0(\Util) \Util
	\modenergy ^{\betagammabig}
	\sum_{l=-\infty}^{+\infty} \Qlm ( \alpha^\prime )
	\Qlm^* ( \alpha ) \Flm (\alpha-\alpha^\prime)
\label{eq:Sm2}.
\end{equation}
Here, to make this expression a little more manageable, we have defined
the integral over angular momentum to be the {\em angular momentum
function} $\Flm(\eta)$, where $\eta = \alpha - \alpha^\prime$:
\begin{equation}
 	 \Flm (\eta)
 = 
	{ 1 \over {2\pi} } \int_0^\infty
	{ {e^{-i \eta {2\over {2-\beta}}  \ln \Ltil }}
	 \over { \lkapmom -\omtil \Ltil^{\twopm}} }
\left[
	\left\{ ( \lkapmom ) \modterm - \gamma m \right\}
	 \Htil ( \Ltil)
	- m \Ltil \frac{d\Htil}{d\Ltil}
\right] { {d\Ltil}  \over \Ltil}.
	\label{eq:Flm}
\end{equation}
For convenience, the Fourier coefficients $\Qlm$ and the angular
momentum function $\Flm$ are shown as depending only upon the
wavenumber $\alpha$ in~\eqref{eq:Sm2}, although they also depend upon
the eccentric velocity $\Util$ as well as on the disk parameters.
The corresponding results for the transfer function and the 
angular momentum function for the Toomre-Zang disk are (Z3.40):
\begin{equation}
\Sm ( \alpha, \alpha^\prime ) = 
	\frac
	{(\gamma+1)^{1+\gamma/2}K(\alpha^\prime,m)  }
	{
	2^{\gamma/2} e^{(\gamma+1)/2} \sqrt{\pi} \Gamma \left[ \frac{\gamma+1}{2} \right]
	}
	\left( \frac{2}{\gamma+1} \right)^{\gamma/2}
	\int d\Util \,
	\auxint_0(\Util) \Util
	e^{-\frac{(\gamma+1)\Util^2}{2} }
	\sum_{l=-\infty}^{+\infty} \Qlm ( \alpha^\prime )
	\Qlm^* ( \alpha ) \Flm (\alpha - \alpha^\prime ) 
  \label{eq:SmMestel}.
\end{equation}
\begin{equation}
	\Flm = { 1 \over {2\pi} } \int_0^\infty
	{ { e^{- i \eta \ln \Ltil}} 
	\over { \lkapmom -\omtil \Ltil} }
\left[
	\left\{
	( \lkapmom ) ( 1 + \gamma ) - \gamma m
	\right\}
	 \Htil ( \Ltil)
	- m \Ltil  \frac{d\Htil}{d\Ltil}
\right] { {d\Ltil}  \over \Ltil}.
\end{equation}
Can we gain some rather more intuitive understanding of this
complicated expression for $\Sm$?  The transfer function $\Sm (
\alpha, \alpha^\prime )$ describes the contribution of the imposed
logarithmic spiral component with wavenumber $\alpha^\prime$ to the
response component with wavenumber $\alpha$.  To see how this is
calculated, we consider a star orbiting in the disk. The shape of its
orbit is characterised by its eccentric velocity $\Util$. The Fourier
coefficient $\Qlm$ describes the ``match'' between a particular
logarithmic spiral component and this orbit. Specifically, the changes
in the star's energy and angular momentum caused by the logarithmic
spiral perturbation are expanded in harmonics of the orbit's radial
period. The Fourier coefficient $\Qlm$ gives the contribution to the
$l$th component of the perturbation experienced by the star.  Thus, in
the expression for the transfer function~\eqref{eq:Sm2}, the first
Fourier coefficient $\Qlm (\alpha^\prime)$ describes how far the star
is forced out of its unperturbed orbit by the imposed perturbation,
and hence the star's tendency to stop contributing to the imposed
logarithmic spiral.  The second Fourier coefficient $\Qlm^* (\alpha)$
describes how well matched the perturbed star is to the response
logarithmic spiral component with wavenumber $\alpha$.  Both these
depend only on the shape of the star's orbit, not on its size; nor has
any consideration yet been made of the ``interaction'' between the
logarithmic spirals.  This is accounted for by the angular momentum
function $\Flm$.  How easy it is for a star to move from the imposed
to the response logarithmic spiral depends on the difference between
the response and imposed wavenumbers, $\eta = \alpha - \alpha^\prime$,
as well as the charactistics of the perturbation (its azimuthal
symmetry, growth rate and pattern speed, given by $m$ and $\omtil$),
and also on the size of the orbit, $\Ltil$. The integral in the
angular momentum adds up similar-shaped orbits of all different
sizes. $\Flm$ then describes how feasible it is for density to move
from wavenumber $\alpha^\prime$ to $\alpha$ in a perturbation with
this $m$ and $\omtil$.  Owing to the scale-free nature of the disk, we
have been able to deal simultaneously with all orbits of a given
shape, irrespective of their size. The $\Util$-dependent parts of the
integrand in $\Sm$ measure how many stars there are for each shape of
orbit. These are then added up to determine how much density
ultimately moves from the logarithmic spiral component with wavenumber
$\alpha^\prime$ to that with wavenumber $\alpha$.


\subsection{The Angular Momentum Function}

\begin{figure}
	\begin{center}
		\epsfig{file=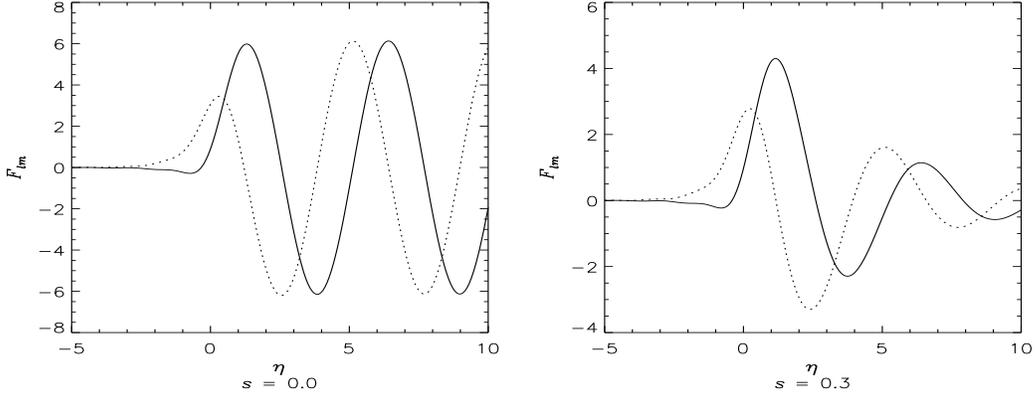,width=0.8\textwidth,height=0.3\textwidth}
		\caption [The angular momentum function for two
		different growth rates, $\beta=0.25$] {Graph of the
		angular momentum function $F_{22}$ plotted against
		$\eta$ for two different growth rates in a
		$\beta=0.25$ disk. In the left panel, $s=10^{-7}$; on
		the right, $s=0.3$. In each case, the solid line is
		the real part, and the dotted line the
		imaginary. [Numerical details: $\beta=0.25$, $N = 2$,
		$M = 6$, $\Ltilc = 31.6$; $\gamma = 13.7$,
		$\Omegap=0.5$, $m = 2$; $l = 2$, $\Util = 0.5$.]
\label{fig:F22_b0p25_U0p5_IO26}
}
	\end{center} 
\end{figure}

\begin{figure}
	\begin{center}
		\epsfig{file=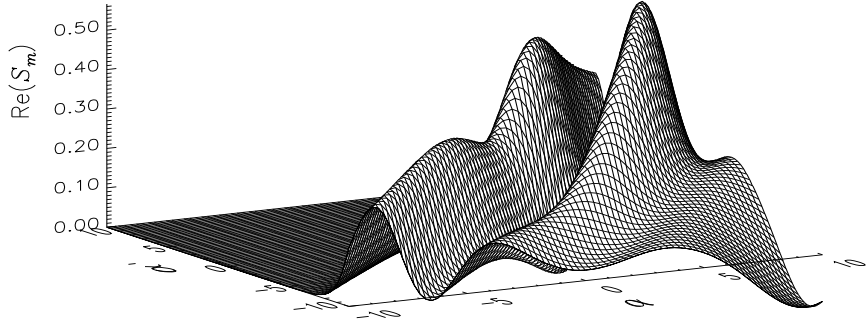,width=0.7\textwidth,height=0.3\textwidth}
		\caption[Trailing bias of the transfer function]{Plot
		of the real part of the transfer function against
		$\alpha$ and $\alpha^\prime$. Notice the
		abrupt change on crossing the line $\alpha =
		\alpha^\prime$ -- this is the trailing
		bias. [Numerical details: $\beta = 0.25$, $N = 2$, $M
		= 6$, $\Ltilc = 31.6$; $\gamma = 13.7$, $\Omegap =
		0.5$, $s=10^{-7}$, $m=2$.]
\label{fig:SmGrid}}
	\end{center} 
\end{figure}

Our job is not quite yet complete! The angular momentum function
eq.~\eqref{eq:Flm} can be worked out analytically for the cut-out
functions $\Htil (\Ltil)$ given in section~\ref{sec:cutouts}. This
contour integration is by no means trivial and it seems wise to
relegate the details of the calculation to Appendix
~\ref{sec:FlmIntegration}.  In this section, we aim to understand the
behaviour of the analytic expression for the angular momentum function
in physical terms.  Fig.~\ref{fig:F22_b0p25_U0p5_IO26} compares the
angular momentum function for two different growth rates $s$. The
left-hand plot shows $\Flm(\eta)$ for vanishing growth rate, while the
right-hand one has $s=0.3$.
This figure exemplifies the trailing bias of these disks. Where
$\lkapmom$ is positive and the eccentric velocity is low, the angular
momentum function is highly asymmetric about $\eta=0$: the decay for
$\eta > 0$ is greatly attenuated. For negative values of $\lkapmom$ or
for high eccentric velocities, $\Flm$ decays rapidly as $|\eta|$ moves
away from zero in either direction.  The mathematical origin of this
asymmetry is seen most simply in the expression for $\Flm$ in the case
of a scale-free disk~\eqref{eq:FlmSingetanonzero}. This depends on
$\eta$ as
\be
	\Flm (\eta) \varpropto
	\frac
	{  e^{-i \hat{\eta} \ln {\lkapmom \over \omtil} } }
	 {1- e^{2\pi\hat{\eta}} },
\end{equation}
where $\hat{\eta}$ is defined as $2\eta /(2+\beta)$ (see
Appendix~\ref{sec:FlmIntegration},~\eqref{eq:hateta}).  If the growth
rate is zero, then on taking the principal value of the logarithm,
we obtain
\begin{xalignat}{2}
	\Flm (\eta) & \varpropto
	e^{-i \hat{\eta} \ln \big|{\lkapmom \over {m\Omegap}} \big| }
	\frac
	{  e^{2\pi\hat{\eta}} }
	 {1- e^{2\pi\hat{\eta}} },
&\quad
	\lkapmom &> 0;
\\
	\Flm (\eta) & \varpropto
	e^{-i \hat{\eta} \ln \big|{\lkapmom \over {m\Omegap}} \big| }
	\frac
	{  e^{\pi\hat{\eta}} }
	 {1- e^{2\pi\hat{\eta}} },
&\quad
	\lkapmom &< 0.
\end{xalignat}
Considering extreme values of $\eta$ on either side of zero, it is
apparent that for $\lkapmom>0$ the angular momentum function is be
asymmetric about $\eta =0$. The magnitude of $\Flm$ tends to a
constant value for large positive $\eta$, whereas it decays rapidly as
$\exp(-2\pi|\hat{\eta}|)$ for negative $\eta$. Conversely, for
$\lkapmom<0$ no such asymmetry is apparent; $\Flm$ decays as
$\exp(-\pi|\hat{\eta}|)$ in either direction.  After the summation
over radial harmonics and integration over eccentric velocity, the
magnitude of $\Flm$ is likely, on average, to be greater for positive
$\eta$ ($\alpha > \alpha^\prime$) than for negative.  This in turn
means that, typically, the magnitude of the transfer function $\Sm (
\alpha, \alpha^\prime)$ is greater for $\alpha > \alpha^\prime$, as
demonstrated in fig.~\ref{fig:SmGrid}.  Most of the response at
$\alpha$ is due to imposed components with wavenumber {\em less} than
$\alpha$. An alternative way of viewing the situation is that imposed
components mostly go to make up response components with {\em larger}
wavenumber than their own.  Thus the asymmetry in the transfer
function means that our disk tends to make whatever pattern is imposed
on it more trailing.

\subsection{The Special Case of Axisymmetry}\label{sec:axisymminteqn}
In the case of an axisymmetric perturbation, the integral equation
admits certain symmetries.  The first simplification is that the
frequency is purely imaginary: $\omega=is$. Secondly, the integral
equation is equivalent to one with a Hermitian kernel, and hence must
have purely real eigenvalues.  It is straightforward to deduce the
symmetries
\begin{equation}
	F_{l0} (\eta)  = F_{-l0}^* (-\eta),
\qquad\qquad
	Q_{l0}^* (\alpha) = Q_{l0} (-\alpha),
\qquad\qquad
	Q_{-l0} (\alpha) = Q_{l0} (\alpha).
	\label{eq:Sym}
\end{equation}
Let us now recast the integral equation~\eqref{eq:eigeninteqn} into a
form with a Hermitian kernel by defining a modified transfer function
and density transform~\cite{Zang:1976}
\begin{equation}
	T_0(\alpha,\alpha^\prime) 
	= \sqrt{\frac{K(\alpha,0)}{K(\alpha^\prime,0)}} 
	\mathcal{S}_0 (\alpha,\alpha^\prime)
\qquad\qquad
	B(\alpha) = \sqrt{K(\alpha,0)} A(\alpha).
	\label{eq:036}
\end{equation}
The integral equation~\eqref{eq:eigeninteqn} now becomes
\be 	
	\lambda B ( \alpha ) 
	=	
	\int_{-\infty}^{+\infty} d\alpha^\prime
	B ( \alpha^\prime ) T_0 ( \alpha, \alpha^\prime ) 
	\label{eq:inteqnm0}.
\end{equation}
The value of this transformation is that $T_0 ( \alpha, \alpha^\prime
) $ is Hermitian, viz
\begin{equation}
	T_0^* ( \alpha^\prime, \alpha )  = T_0 ( \alpha, \alpha^\prime ).
\end{equation}
The modified integral equation~\eqref{eq:inteqnm0} is thus a
homogeneous, linear Fredholm integral equation with a Hermitian
kernel. It must therefore have real eigenvalues and orthogonal
eigenfunctions~\cite{Tricomi:1985}.  This occurs because the
mathematical eigenvalues have no dependence on pattern speed.  In
general, the mathematical eigenvalue $\lambda$ is an analytic function
of the complex frequency $\omega$. Only for $m=0$ is the frequency
purely imaginary and the problem one-dimensional.

There is a further symmetry. We make the transformations $\alpha
\rightarrow -\alpha$, $\alpha^\prime \rightarrow -\alpha^\prime$, $l
\rightarrow -l$, so as to obtain $T_0^* ( -\alpha,
-\alpha^\prime ) $.  Using the symmetry properties of the Fourier
coefficients and the angular momentum function~\eqref{eq:Sym} along
with that of the Kalnajs gravity factor, we obtain the result
\begin{equation}
	T_0^* ( -\alpha, -\alpha^\prime )  = T_0 ( \alpha, \alpha^\prime ).
\end{equation}
This has the consequence that the eigenvalues are degenerate. If $B
(\alpha)$ is an eigenvector with eigenvalue $\lambda$, then so is $B^*
(-\alpha)$. To see this, we make the transformations $\alpha
\rightarrow -\alpha$, $\alpha^\prime \rightarrow -\alpha^\prime$ in
the modified integral equation~\eqref{eq:inteqnm0}, and take the
complex conjugate. We obtain:
\be 	
	\lambda B^* (- \alpha ) 
	= 
	-\int_{\infty}^{-\infty} d\alpha^\prime
	B^* (- \alpha^\prime ) 
	 T_0^* ( -\alpha, -\alpha^\prime )
	= 
	\int_{-\infty}^{+\infty} d\alpha^\prime
	B^* (- \alpha^\prime ) 
	 T_0 ( \alpha, \alpha^\prime ).
\end{equation}
In other words, $B^* (-\alpha)$ also satisfies the integral
equation~\eqref{eq:inteqnm0}.  In terms of our original density
transforms, $A (\alpha)$ and $A^* (-\alpha)$ are a degenerate pair
with the same eigenvalue $\lambda$.  This is just the anti-spiral
theorem~\cite{LBOst:1967,Kalnajs:1971} for axisymmetric perturbations.


\section{Numerical methods}\label{sec:ChapNum}

This section discusses the numerical algorithms required to find the
normal modes. Sections 4.1 and 4.2 consider the evaluation of the
Fourier coefficients and the transfer function respectively, whereas
Section 4.3 presents the discretisation of the integral
equation. Again, let us emphasise that the numerical method is adopted
-- with a little streamlining -- from Zang~\shortcite{Zang:1976}.


\subsection{The Fourier Coefficient}~\label{sec:Qlm}
\begin{figure}
	\begin{center}
		\epsfig{file=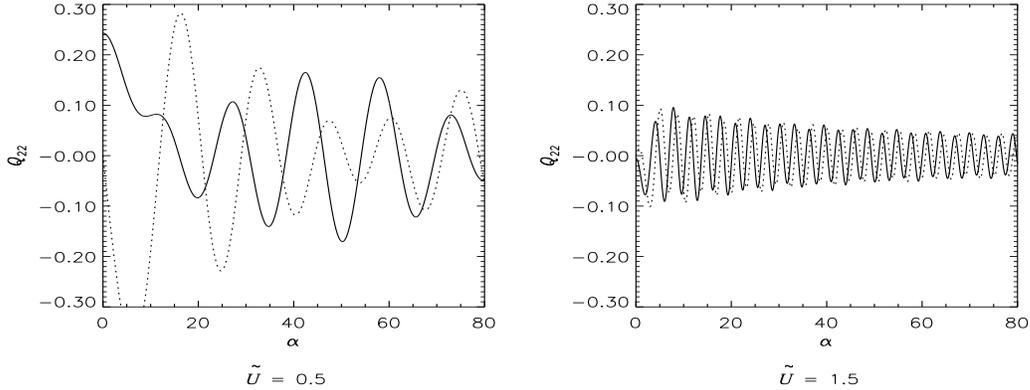,width=0.8\textwidth,height=0.3\textwidth}
		\caption [The Fourier coefficient for two different
		eccentric velocities] {Graph of the Fourier
		coefficient $Q_{22}$ plotted against wavenumber
		$\alpha$. In the left-hand plot, $\Util=0.5$; on the
		right, $\Util=1.5$. In each case, the solid line is
		the real, and the dotted line the imaginary, part of
		$Q_{22}$. [Numerical details: $\beta=0.25$, $m=2$,
		$l=2$; $a_{\text{acc}}=100$, $b_{\text{acc}}=2.5$.]
		\label{fig:Q22_b0p25}} \end{center}
\end{figure}

\begin{figure}
	\begin{center}
		\epsfig{file=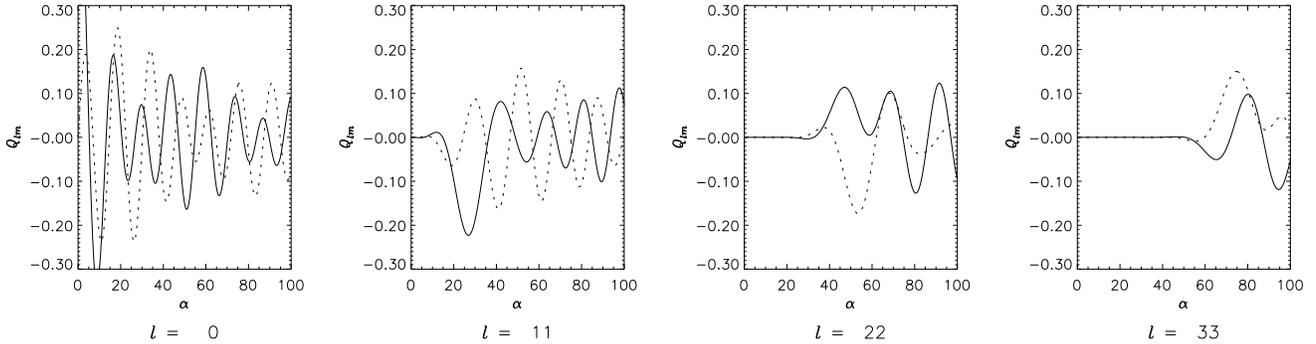,width=\textwidth,height=0.25\textwidth}
		\caption [The Fourier coefficient for four different
		radial harmonics $l$] {Graph of the Fourier
		coefficient $\Qlm$ plotted against wavenumber $\alpha$
		for four different radial harmonics $l$. In each case,
		the solid line is the real, and the dotted line the
		imaginary, part of $\Qlm$. [Numerical details:
		$\beta=0.25$, $m=2$, $\Util=0.5$;
		$a_{\text{acc}}=100$, $b_{\text{acc}}=2.5$.]
		\label{fig:Qmanyl_b0p25}} \end{center}
\end{figure}

\noindent
The integrand in the definition of the Fourier
coeffients~\eqref{eq:Qlm} is periodic with period $2\pi$ in $\chi$.
We are free to define $\chi=0$ to correspond to pericentre. This means
that at the time $t=0$, the star has radial coordinate $R=R_{\min}$.
As $\Xtil$ is even, and $\Ytil$ is odd, about $\chi = 0$, we can
write
\begin{equation}
 	\Qlm(\alpha) ={1 \over \pi} 
		\int_0^\pi
		\exp\left\{ \big( i  \alpha - \fr12 \big) \Xtil \right\}
		\cos ( m \Ytil -  l \chi ) \,	d\chi	.
	\label{eq:Qlmcos}
\end{equation}
The equations of motion~\eqref{eq:dimeqmot} are solved by fourth-order
Runge-Kutta integration~\cite[chap. 15]{NumRec} to obtain the stellar
position as a function of time, and thus $\Xtil$ and $\Ytil$ as a
function of $\chi$. The integration over $\chi$ is carried out by the
midpoint method~\cite[chap. 4]{NumRec}, using $n_{\psi}$ points in
the midpoint integration and $2 n_{\psi}$ in the Runge-Kutta. The
problem is to choose $n_{\psi}$ large enough to obtain excellent
accuracy, while keeping it as small as possible in order to save
time. Eccentric orbits need more work to obtain good accuracy, so, as
suggested by Zang~\cite*{Zang:1976}, $n_\psi$ is made to depend
exponentially on $\Util$:
\be
	n_\psi = a_{\text{acc}} \exp ( b_{\text{acc}} \Util ).
	\label{eq:npsi}
\end{equation}
Values of $a_{\text{acc}} = 10$ and $b_{\text{acc}} = 1.5$ usually
worked well. With these values, the Fourier coefficient is obtained
with around 6 s.f. accuracy for low eccentricity orbits ($\Util =
0.5$), but perhaps only 1 or 2 s.f. for high eccentricity orbits like
($\Util = 1.5$). The Fourier coefficients are generally smaller for
higher values of the eccentric velocity. In the integrand for the
transfer function, two Fourier coefficients are multiplied together at
every eccentric velocity. The accuracy with which the transfer
function is obtained is thus dominated by the accuracy of the Fourier
coefficients at low eccentric velocities. This is useful, since the
Fourier coefficients are costly to evaluate when the eccentric
velocity is high.  Fig.~\ref{fig:Q22_b0p25} shows the Fourier
coefficient as a function of wavenumber, for two different eccentric
velocities. It is shown only for positive wavenumber. Its behaviour
for negative wavenumber is readily deduced from the symmetry property
$\Qlm (\alpha ) = \Qlm^* (-\alpha)$.  We see that $\Util$ controls the
amplitude and the frequency of oscillation of $\Qlm$.
Fig.~\ref{fig:Qmanyl_b0p25} compares Fourier coefficients at different
radial harmonics. For high values of $l$, $\Qlm$ remains close to zero
until $\alpha$ is large; thereafter it oscillates with lower
frequency. Logarithmic spiral components must be tightly-wound in order
to excite responses at high radial harmonics.
 

\subsection{The Transfer Function}
The expression for the transfer function~\eqref{eq:Sm2} involves a sum
over radial harmonic $l$ from $-\infty$ to $+\infty$.  Fortunately,
the magnitude of the terms in this sum decreases sharply with
$|l|$. Negative values of $l$ decrease their contribution faster than
positive $l$. A good approximation is given by summing from $l =
l_{\min}$ to $l = l_{\max}$, where $l_{\min}$ is negative and
$|l_{\min}|$ is less than $l_{\max}$. Values of $l_{\min} = -20$,
$l_{\max} = 30$ usually worked well.  The transfer function also
contains an integral over eccentric velocity $\Util$, which adds up
orbits of all possible shapes.  The equivalent expression for the
Toomre-Zang disk~\eqref{eq:SmMestel} contains a Gaussian factor $\exp
(-\half \Util^2/ \sigutil^2)$. This prompted Zang~\cite*{Zang:1976} to
use Gauss-Laguerre quadrature to evaluate $\Sm$. The fundamental
formula is
\be
	\int_0^\infty f(x) e^{-x} dx \approx \sum_{i=1}^{n_{GL}} w_i f(x_i)
	\label{eq:GL},
\end{equation}
where $f(x)$ is a smooth polynomial-like function.  The weights $w_i$
and abscissae $x_i$ are well-known (see Abramowitz \& Stegun 1989;
Press {\em et al.} 1989, chap. 4). Zang~\cite*{Zang:1976} found it
preferable to reduce the dispersion of the Gaussian in this
expression, replacing it with $\tilde{\sigma}_n = f_\sigma \sigutil$,
where the fraction $f_\sigma$ is around 80\%. This concentrates
attention on the lower values of $\Util$, where the integrand is
larger.  For the power-law disks, the distribution of radial
velocities at any spot is similar -- but not exactly equal to -- a
Gaussian with dispersion $\sigutil$. This suggests that Gauss-Laguerre
quadrature may still work well. Although our integrand is not directly
of form~\eqref{eq:GL}, it is readily made so. The limits on our
integral are zero and infinity for negative $\beta$. For positive
$\beta$, the upper limit is $\Util=( 2/\beta - 1)^{1/2}$.  At least
for the models of most interest in galactic astronomy ($|\beta|
\lesssim 0.5$), this distinction is in practice unimportant, since the
integrand has fallen to negligibly small values well before this limit
on the eccentric velocity.

Let $\mathcal{I}_U$ be the integrand in $\Sm$ when the integration is
carried out over $\Util$, so $\Sm \left( \alpha, \alpha^\prime \right)
\varpropto \int_0^{\infty} \mathcal{I}_U d\Util $. We then have
\be
	\mathcal{I}_V = \mathcal{I}_U \frac{\tilde{\sigma}_n^2}{\Util} 
	\exp \left( \frac{\Util^2}{2 \tilde{\sigma}_n^2} \right),
\end{equation}
where $\Sm \left( \alpha, \alpha^\prime \right) \varpropto
\int_0^{\infty} \mathcal{I}_V e^{-V}dV $ and $V = \half\Util^2 /
\tilde{\sigma}_n^2$. The reliability of Gauss-Laguerre quadrature 
was tested against the extended midpoint method~\cite[eq. (4.1.19)]{NumRec}. 
These comparisons~\cite{Read:1997} indicate that Gauss-Laguerre
quadrature is a remarkably efficient way of performing the
integral. Excellent six significant figure accuracy is obtained with a
handful of function evaluations.


\subsection{Computational Solution of the Integral Equation}\label{sec:compsoln}

\noindent
Given that the kernel of the integral equation can be evaluated to
high accuracy, we now need to devise a method for its numerical
solution. We ultimately seek self-consistent solutions, for which
$\Ares (\alpha) = \Aimp (\alpha )$. However, as Zang (1976) already
noted, the first step is to consider the more general problem for
which the response density is a complex multiple of the imposed
density, i.e., $\Ares (\alpha) =
\lambda \Aimp (\alpha )$. This casts the integral equation~\eqref{eq:inteqn} 
into the form
\be 	
	\lambda A \left( \alpha \right) =	\int_{-\infty}^{+\infty} d\alpha^\prime
	A \left( \alpha^\prime \right) \Sm ( \alpha, \alpha^\prime ) 
	\label{eq:eigeninteqn},
\end{equation}
where we have dropped the subscripts on $A$. We refer to $\lambda$ as
the mathematical eigenvalue. Of course, only the instance when
$\lambda$ is unity carries the physical significance of a mode. The
advantage of this mathematical artifice is that the integral
equation~\eqref{eq:eigeninteqn} is in the standard form of a
homogeneous, linear, Fredholm equation of the second kind (see
e.g. Courant \& Hilbert 1953; Delves \& Mohamed 1985). For a given
value of $\lambda$, such an equation normally admits only the trivial
solution, $A(\alpha) \equiv 0$. The values of $\lambda$ for which
non-trivial solutions exist are the eigenvalues of the equation. We
seek an iterative scheme which drives the eigenvalue to unity, thus
providing a self-consistent mode.  There is a close analogy between
linear algebraic equations and linear integral equations. The former
define relations between vectors in a finite-dimensional vector space,
the latter define relations between functions in an
infinite-dimensional vector space (technically, a Banach space). This
analogy can be made explicit by applying a quadrature rule to the
integration in~\eqref{eq:eigeninteqn} to obtain
\be 	
	\lambda A \left( \alpha_j \right) =	\sum_i w_i
	A \left( \alpha_i \right) \Sm ( \alpha_j, \alpha_i ) ,
\end{equation}
where $w_i$ are some appropriate weights. This general approach is
called the Nystrom method~\cite{Del:Moh}.  It evidently reduces the
solution of an integral equation to the solution of an algebraic
eigenvalue problem. The latter is a classic and well-studied area of
numerical analysis, for which many tried and tested techniques are
available.

Much of the skill in the numerical solution of integral equations
comes from the choice of quadrature rules and weights.
Zang~\cite*{Zang:1976} devised an elegant method based on locally
approximating the kernel and the response by Lagrangian interpolating
polynomials. This is naturally adapted to the instance when the kernel
varies on a much smaller scale than the solution. We follow this
method, but we did not find Zang's splitting of the kernel of the
integral equation into Hermitian and Volterra parts to be needed.
First, to obtain smoother functions, the Kalnajs gravity factor is
extracted by defining
\begin{xalignat}{2}
	\Sm \left( \alpha, \alpha^\prime \right) 
		& = K\left(\alpha^\prime,m\right) \Smtil \left( \alpha, \alpha^\prime \right),
	\label{eq:smoothSm}
& \quad
	A \left( \alpha \right) &= \frac{\Atil (\alpha)}{ K\left(\alpha,m\right) }.
	\label{eq:smoothA}
\end{xalignat}
The eigenvalue equation now becomes
\be 	
	\lambda \Atil \left( \alpha\right) = K \left( \alpha,m \right) 
	\int_{-\infty}^{+\infty} d\alpha^\prime
	\Atil \left( \alpha^\prime \right) \Smtil \left( \alpha, \alpha^\prime \right)
\label{eq:inteqntil}.
\end{equation}
Let us introduce a finite grid of points in wavenumber space,
${\alpha_r}$. If we need to know the value of $\Atil$ or $\Smtil$ at a
value of $\alpha$ intermediate between the gridpoints $\alpha_r$ and
$\alpha_{r+1}$, we interpolate over the eight gridpoints from
$\alpha_{r-3}$ to $\alpha_{r+4}$.  For 8 equally-spaced points $\Delta
\alpha$ apart, Lagrange's classic formula for the interpolating
polynomial $P(\alpha)$ through $N$ points $f(\alpha_k)$
~\cite[chap. 3]{NumRec} becomes
\begin{equation}
	P(\alpha) = \prod_{i=1}^{8} \left( \alpha - \alpha_{r+i-4} \right)
		\sum_{k=-3}^{4} 
		\frac{(-1)^k }{(3+k)! (4-k)! }
		\frac{ f(\alpha_{r+k})}
		{ (\Delta\alpha)^7 (\alpha - \alpha_r - k \Delta \alpha)}.
\end{equation}
Defining $x = ( \alpha-\alpha_r ) /\Delta\alpha$, this becomes
\begin{equation}
	P(\alpha) = \prod_{i=1}^{8} \left( x + 4 - i \right)
		\sum_{k=-3}^{4} (-1)^k 
		\frac{f(\alpha_{r+k})}
		{(3+k)! (4-k)! (x-k)}
	= \sum_{k=-3}^{4} L_k [x] f(\alpha_{r+k}),
\end{equation}
where
\be
	L_k [x] = (-1)^k
		\frac{\prod_{i=1}^{8} \left( x + 4 - i \right)}
		{(3+k)! (4-k)! (x-k)}.
\end{equation}
Then the interpolated approximations to the response and the transfer
function are (cf. Zang 1976, app. D)
\begin{xalignat}{2}
	\Atil (\alpha^\prime) 
	& \approx \sum_{k=-3}^{4} L_k [ x^\prime ]
	\Atil (\alpha_{r+k} )
\label{eq:Atilinterp},
	& \quad
	\Smtil (\alpha, \alpha^\prime) 
	& \approx
	\sum_{k=-3}^{4} L_k [ x^\prime ]
	\Smtil (\alpha, \alpha_{r+k} ) 
\label{eq:Smtilinterp},
\end{xalignat}
where $x^\prime = ( \alpha^\prime-\alpha_r ) /\Delta\alpha$.  The
infinite range in the integral equation~\eqref{eq:eigeninteqn} is a
kind of singularity. Truncation of the wavenumber range at large
finite values is the simplest but most brutal way of handling the
singularity. In practice, we found this to be surprisingly
effective. Wavenumber space is approximated by a finite grid with $n$
points along each side.  We choose $n$, and the grid-point spacing
$\Delta \alpha$, so as to obtain a sufficiently accurate solution of
the integral equation~\eqref{eq:inteqn}.  The finite size of the grid
means that we encounter problems when evaluating $\Atilimp$ and
$\Smtil$ for values of $\alpha$ near the edges of the grid. We are
interpolating over eight points, so we need data from $\alpha_{-2}$ to
$\alpha_{n+3}$. We deal with this problem by simply assuming that our
function is zero outside the grid.  This is acceptable provided the
grid is large enough. Then the values of the kernel at the missing
interpolation points are negligibly small. This procedure is justified
empirically by the demonstration that the mathematical eigenvalue
converges to a value independent of grid size~\cite{Read:1997}.

Substituting the Lagrange-interpolated approximations for $\Atilimp$
and $\Smtil$ ~\eqref{eq:Atilinterp} into the modified integral
equation~\eqref{eq:inteqntil}, we obtain
\begin{equation} 	
	\lambda \Atil \left( \alpha\right) = K \left( \alpha,m \right) 
	\sum_{i=-3}^{4} \sum_{k=-3}^{4}
	\int_{\alpha_1}^{\alpha_n} d\alpha^\prime
	L_i \left[ \frac{\alpha^\prime-\alpha_r}{\Delta\alpha}\right] 
	L_k \left[ \frac{\alpha^\prime-\alpha_r}{\Delta\alpha}\right]
	\Atil (\alpha^\prime_{r+i} )
	\Smtil (\alpha, \alpha^\prime_{r+k} ).
	\label{eq:012}
\end{equation}
Of course, the integration over wavenumber now runs from $\alpha =
\alpha_1$ to $\alpha_n$, instead of from $\alpha = -\infty$ to
$\infty$.  We break this integration into $n$ portions of $\Delta
\alpha$.
\begin{equation} 	
	\lambda  \Atil \left( \alpha\right)
 = 
	K \left( \alpha,m \right) 
	\sum_{r=1}^{n} \sum_{i=-3}^{4} \sum_{k=-3}^{4}
	\int_{\alpha_r}^{\alpha_r + \Delta \alpha} d\alpha^\prime
	L_i \left[\frac{\alpha^\prime-\alpha_r}{\Delta\alpha}\right] 
	L_k \left[\frac{\alpha^\prime-\alpha_r}{\Delta\alpha}\right]
	\Atil (\alpha^\prime_{r+i} )
	\Smtil (\alpha, \alpha^\prime_{r+k} ).
\end{equation}
Then, changing variables to $x^\prime = (\alpha^\prime-\alpha_r) /
\Delta\alpha$, we obtain
\begin{equation}
	\lambda \Atil \left( \alpha\right) = K \left( \alpha,m \right) 
	\sum_{r=1}^{n} \sum_{i=-3}^{4} \sum_{k=-3}^{4}
	\Atil (\alpha_{r+i} )
	\Smtil (\alpha, \alpha _{r+k} )
	\Delta \alpha \int_{0}^{1} dx^\prime L_i [x^\prime] L_k [x^\prime].
\end{equation}
Let us define the weighting coefficients $C_{ik}$ by
(cf. Zang~\cite*{Zang:1976}, app. D)
\be
	 C_{ik} = \Delta \alpha \int_{0}^{1} dx L_i [x] L_k [x].
\end{equation}
Although there appear to be 64 $C_{ik}$, only 20 of them are
independent, because of the symmetry properties $C_{ik} \equiv
C_{ki}$, and $C_{ik} \equiv C_{(1-i)(1-k)}$.  The weighting
coefficients are easily evaluated using the midpoint method given in
Press {\em et al.}~\cite*[chap. 4]{NumRec}. They are tabulated in
Read~\cite*{Read:1997}.  Equation~\eqref{eq:012} can now be written
as:
\be 	
	\lambda \Atil \left( \alpha_j \right) = K \left( \alpha_j,m \right) 
	\sum_{r=1}^{n} \sum_{i=-3}^{4} \sum_{k=-3}^{4}
	C_{ik} \Smtil (\alpha_j, \alpha_{r+k} )
	\Atil (\alpha_{r+i} )
\label{eq:MultSum}
\end{equation}
Terms in this multiple sum for which $r+i$ is less than 1 or greater
than $n$ contribute nothing. We can thus write the above equation in
terms of a single sum from $\alpha_1$ to $\alpha_n$, and collect the
weighting coefficients $C_{ik}$, smoothed transfer function $\Smtil$
and density profile $\Atil$ together into a single quantity
$\mathbf{S}$.
\be 	
	\lambda \Atil (\alpha_j ) = \sum_{s=1}^{n} S_{js} 
		\Atil (\alpha_s ) .
	\label{eq:mateqn}
\end{equation}
This is of course just a matrix equation for the mathematical
eigenvalue $\lambda$.  If there is an eigenvalue of the matrix
$\mathbf{S}$ equal to unity, then the corresponding eigenvector
$\mathbf{A}$ gives us a self-consistent mode, i.e. a self-sustaining
density perturbation.  $S_{js}$ cannot be simply expressed in terms of
$C_{ik}$, $\Smtil$ and $\Atil$. But we see that it represents the
total coefficient of $\Atil (\alpha_s)$ in eq.~\eqref{eq:MultSum}.

Having obtained the matrix $S_{js}$, it is reasonably straightforward
to find its eigenvalues using the eigensystem package ({\it EISPACK})
developed by Smith {\em et al.}~\cite*{EISPACK}. This returns the
eigenvalues of a matrix and, optionally, the corresponding
eigenvectors. The power method can also be used to provide an
independent check on the largest eigenvalue. This relies on repeated
application of the matrix $\mathbf{S}$ to a test vector
$\mathbf{t}$~\cite[chap. 8, sec. 7]{RalRab:1978}.  Probably the most
important factor in achieving excellent accuracy in the eigenvalues is
the size and fineness of the grid.  The range of the grid, $(n-1)
\Delta \alpha$, typically needs to be around 50 in order for the
mathematical eigenvalue to be accurate to 6 s.f. A grid-spacing of
$\Delta\alpha=0.2$ is usually sufficient. However, suitable values
necessarily depend to some extent on the form of the mode. For
instance, for high azimuthal harmonics, the modes tend to be
tightly-wrapped, requiring a more extensive grid.

Having established that the mathematical eigenvalue can be found to
good accuracy, let us consider how to locate the unit eigenvalues
which correspond to self-consistent modes.  We are investigating modes
of a given azimuthal symmetry in a disk with given density
profile. This means that $m$, $\beta$, $N$, $M$ and $\Ltilc$ are held
fixed. It leaves us free to adjust $\gamma$, $s$ and $\Omegap$. A disk
at a given temperature, if it admits a mode at all, generally does so
only for a particular growth rate and pattern speed. We therefore hold
$\gamma$ fixed, and adjust $s$ and $\Omegap$ until a mode is
found. This is in fact a search for the root of the equation $\lambda
=1$ in only one dimension, since the mathematical eigenvalue $\lambda$
is an analytic function of the complex frequency $\omega =
m\Omegap+is$. The Newton-Raphson method is extremely effective at
locating such roots.  When the disk is sufficiently hot, it is
generally completely stable. As the disk is cooled, instabilities set
in through the marginal modes, for which the growth rate $s$ is
zero. We are therefore often interested in the marginally stable
modes. To find these, $s$ is set to some vanishingly small value, and
a two-dimensional search is performed for the critical pattern speed
$\Omegap$ and temperature $\gamma$ using the Newton-Raphson method in
two dimensions.

\section{Conclusions}

\noindent
This paper has set up the machinery for an investigation into the
large-scale, global, modes of the power-law disks. The Fredholm
integral equation for the normal modes has been derived.  This is done
by linearising the collisionless Boltzmann equation to find the
response density corresponding to any imposed density and potential.
The normal modes are given by equating the imposed density to the
response density. This scheme is implemented in Fourier space, so that
the Fredholm integral equation relates the transform of the imposed
density to the transform of the response density. Numerical strategies
are given to discretise the integral equation to yield a matrix
equation. This requires some care as the kernel of the integral
equation is either singular (as in the case of the self-consistent
disk) or almost so (the cut-out disks).  The problem of locating the
normal modes is thus reduced to an algebraic eigenvalue problem, for
which a wealth of standard techniques are available.  The crucial
simplification underlying the analysis is that the force law
describing the gravity field of the galactic disk is scale-free. It is
this that ensures that orbits passing through any one spot in the disk
are simply scaled copies of the orbits passing through any other spot.
When the orbits are forced by a scale-free logarithmic spiral
perturbation, the response of all orbits of the same eccentricity can
be computed at the same time. It only remains to add up the
contributions at every eccentricity. For a disk with an arbitrary
force law, such simplification is not possible.

Let us again emphasise that nearly all the computational techniques
required to study the stability of the power-law disks were developed
by Zang~\shortcite{Zang:1976} in his pioneering study of the disk with
a completely flat rotation curve. Our contribution consists {\it only}
of deriving the extensions to study the complete family of scale-free
power-law disks.  This paper has discussed just the algorithms -- the
aim has been to present all the mathematical and computational details
for reference here.  The following paper in this issue of {\it Monthly
Notices} implements the machinery discussed here to provide a complete
description of the spiral modes of the self-consistent and cut-out
power-law disks.  A fast computer code is developed which gives the
growth rates and pattern speeds of the normal modes for any power-law
disk in a matter of minutes when run on a modern workstation.

\bigskip
\noindent
NWE thanks Alar Toomre for much generous and helpful advice on how to
extend the analysis in Zang's thesis to the family of power-law
disks. NWE is also grateful to the trustees of the Lindemann
Fellowship for an award which enabled him to spend the year of 1994 at
the Massachusetts Institute of Technology. He is supported by the
Royal Society. JCAR thanks the Particle Physics and Astronomy Research
Council for a research studentship.

\bibliographystyle{mnbib}
\bibliography{biblio}

\setcounter{section}{0}
\setcounter{equation}{0}
\setcounter{figure}{0}
\setcounter{table}{0}
\appendix
\section{Single-Eccentricity Distribution Functions}\label{sec:SingEcc}
In the main body of the paper, we have used stellar distribution
functions for the power-law disks that are built from powers of energy
and angular momentum. Many other equilibrium distribution functions
are possible.  In this Appendix, we consider disks in which all the
stars have the same shape of orbit, characterised by an eccentric
velocity $U = U_r$. Mathematically, we look for distribution functions
of the form
\be
	\fsing = F (\RH) \delta ( U - U_r ).
\end{equation}
The surface density is
\be
	\Sigeq = \Sigma_0 \left( \frac{R_0}{R} \right) ^{1+\beta} = \iint \fsing(u,v) du \, dv.
\end{equation}
We transform this to an integral over $U$ and $R_H$, using the Jacobian
\begin{equation}
du\, dv = dU \, dR_H \left( 1 - \frac{\beta}{2} \right)
	\frac{v_\beta}{R}
\left( \frac{R_0}{R_H} \right) ^{\beta/2}
\frac {\Util}{ \sqrt{\Util^2 + 1 - \Rtil^{-2} + \frac{2}{\beta} (\Rtil^{-\beta} - 1) } },
\end{equation}
and substitute our assumed distribution function
\begin{equation}
	\Sigma_0 \left( \frac{R_0}{R} \right) ^{1+\beta}  =
	\left( 1 - \frac{\beta}{2} \right)
	\frac{v_\beta}{R}
	 \iint 
	\left( \frac{R_0}{R_H} \right) ^{\beta/2}
	\frac 
	{\Util \, F (\RH) \, \delta ( U - U_r )  \, dU \, dR_H  }
	{ \sqrt{\Util^2 + 1 - \Rtil^{-2} + \frac{2}{\beta} (\Rtil^{\beta} - 1) } }.
\end{equation}
We integrate over $U$, and then transform the integral over $\RH$ to
one over $\Rtil = R / R_H$:
\begin{equation}
	\Sigma_0 \left( \frac{R_0}{R} \right) ^{1+\beta/2}  =
	\left( 1 - \frac{\beta}{2} \right)
	2 v_\beta \Util_r
	\int_{\Rtilmin}^{\Rtilmax} 
	\frac 
	{ \Rtil^{\beta / 2} F \left( R / \Rtil \right) \, d \Rtil }
	{ \Rtil^2 \sqrt{\Util_r^2 + 1 - \Rtil^{-2} + \frac{2}{\beta} (\Rtil^{\beta} - 1) } },
	\label{eq:031}
\end{equation}
where $\Util_r$ is the value of $U_r$ in dimensionless units.
Looking at the powers of $R$, we can guess at a solution of this
integral equation, namely $ F (R/\Rtil ) = k (\Rtil/R )^{1+\beta/2}.$
Substituting this into~\eqref{eq:031}, and remembering the definition
of the auxiliary integral $\auxint_n ( \Util )$~\eqref{eq:genintI}, we
can solve for $k$ and obtain
\be
	k = 
	\frac
	{\Sigma_0 R_0^{1+\beta/2} }
	{( 1 - \frac{\beta}{2} ) v_\beta \Util_r \auxint_{1-\beta} ( \Util_r )}
 = 
	\frac
	{\Sigma_0 R_0^{1+\beta}  }
	{\left( 1 - \frac{\beta}{2} \right) \RH^{\beta/2} U_r \auxint_{1-\beta} ( \Util_r )}.
\end{equation}
The distribution function is then
\be
	\fsing (U,\RH ) = \frac
	{\Sigma_0 R_0^{1+\beta} }
	{\left( 1 - \frac{\beta}{2} \right) U_r \auxint_{1-\beta} ( \Util_r )}
	\frac{\delta ( U - U_r )}{\RH^{1+\beta}}
	\label{eq:singeccdf}.
\end{equation}
This expression is valid for $\beta=0$, in which case the appropriate
form for the auxiliary integral $\auxint_1$~\eqref{eq:genintIbzero}
must be used.

All the stars in this disk have orbits of the same shape, but
different size. Each star sweeps out an annulus as it orbits. The
relative width of this annulus depends on the eccentric velocity
$U_r$, but the overall size of the annulus depends on $\RH$. It is
easy to imagine that if we add up annuli with a variety of $\RH$ in
suitable proportions, we could recover the surface density of the
equilibrium disk; and this is what the distribution
function~\eqref{eq:singeccdf} does.  It seems intuitively right that
the number of annuli needed falls off with $\RH$ at the same rate as
the surface density falls off with $R$. Similarly, we can see why the
total number of annuli depends inversely on eccentric velocity $U_r$:
for high $U_r$, the annuli are very wide, and few are needed; for low
$U_r$, the converse is true.

\setcounter{equation}{0}
\setcounter{figure}{0}
\setcounter{table}{0}
\appendix
\section{Reference Table of Dimensionless Quantities}\label{sec:RefTabs}


\label{tabl:dimless}
\begin{tabular}%
{|l|l|l|l|}
\hline
& Quantity
& For the Toomre-Zang disk
& For the general power-law disk
\\ \hline
$\RtilH$
& Home radius
& $\Rtil_H = {{\RH}/{R_0}}$
& $\Rtil_H = {{\RH}/{R_0}}$
\\ 
$\tilde{t}$ 		
& Time 	
& $\tilde{t} = {{v_0} \over {\RH}} t$		
& $\tilde{t} = {{v_\beta} \over {\RH}} \Rtil_H^{-\beta/2} t$	
\\ 
$\tilde{u}$ 		
& Radial velocity
& $\tilde{u} = { {d \Rtil} \over {d \tilde{t}}} = {u \over v_0} $ 
& $\tilde{u} = { {d \Rtil} \over {d \tilde{t}}} = {u \over v_\beta} \Rtil_H ^{\beta / 2} $
\\ 
$\tilde{v}$	
& Tangential velocity
& $\tilde{v} = { {\Rtil d \theta } \over {d \tilde{t}}}= {v \over v_0} $ 
  $\tilde{v} = \Rtil^{-1} $
& $\tilde{v} = { {\Rtil d \theta } \over {d \tilde{t}}}= {v \over v_\beta} \Rtil_H^{\beta / 2} $
  $ \tilde{v} = \Rtil^{-1}$
\\ 
$\Etil$
& Energy 
& $ \Etil = {E \over {v_0^2} }$ ;
  $\Etil= \half \left( \tilde{u}^2 + \tilde{v}^2 \right) - \ln \frac{R_0}{R}$
& $ \Etil  = {E \over v_\beta^2} \Rtil_H ^\beta$ ;
  $\Etil= \half \left( \tilde{u}^2 + \tilde{v}^2 \right) - {1 \over {\beta \Rtil^\beta}}$
\\ 
$\Ltil$
& Angular momentum
& $\Ltil = {L_z \over {R_0 v_0} } = \Rtil$
& $\Ltil = {L_z \over {R_0 v_\beta} } = \Rtil^{1-\beta/2}$
\\ 
$\Util$
& Eccentric velocity
& $\Util  = {U \over v_0} $ 
& $\Util  = {U \over v_\beta} \Rtil_H ^{\beta / 2} $ 
\\ 
&
& $\Util^2 = \tilde{u}^2 - 1 + \Rtil^{-2} + 2 \ln{\Rtil}$
& $\Util^2 = \tilde{u}^2 - 1 + \Rtil^{-2} - {2 \over \beta} \left( {\Rtil^{-\beta} -1 } \right)$
\\ 
$\auxint_n$
& Auxiliary integral
& $ \auxint_n  = \int_{\Rtilmin}^{\Rtilmax} {{2d\Rtil}
\over {\Rtil^n  \left( \Util^2 + 1 - \Rtil^{-2} -2 \ln \Rtil  \right) ^\half }}$
& $ \auxint_n  = \int_{\Rtilmin}^{\Rtilmax} {{2d\Rtil}
\over {\Rtil^n \left( \Util^2 + 1 - \Rtil^{-2} 
+ {2 \over \beta} \left( \Rtil^{-\beta} - 1 \right) \right) ^\half }}$
\\ 
$ \kaptil$
& Radial frequency
& $\kaptil = \frac{\RH}{v_0} \kappa 
	= {{2\pi}\over {\auxint_0}}$
& $\kaptil = \frac{\RH}{v_\beta} \Rtil_H^{\beta/2} \kappa
	= {{2\pi}\over {\auxint_0}}$
\\ 
$ \Omtil$
& Angular frequency
& $\Omtil =  \frac{\RH}{v_0} \Omega = {{\auxint_2}\over {\auxint_0}}$
& $\Omtil =  \frac{\RH}{v_\beta} \Rtil_H^{\beta/2} \Omega
	= {{\auxint_2}\over {\auxint_0}}$
\\ 
$\kaptil_0$
& Epicyclic frequency
& $\kaptil_0  = \sqrt{2}$
& $\kaptil_0 = \sqrt{2-\beta}$
\\ 
$\Omtil_0$
& Circular frequency
& $\Omtil_0 = 1 $
& $\Omtil_0 = 1$
\\ 
$\chi$
& Orbital phase
& $\chi = \kappa t = \kaptil \tilde{t}$
& $\chi = \kappa t = \kaptil \tilde{t}$
\\ 
$\Ytil$
& Angular deviation
& $\Ytil = \theta - \Omega t = \theta - \Omtil \tilde{t}$
& $\Ytil = \theta - \Omega t = \theta - \Omtil \tilde{t}$
\\ 
$\xtil$
& Logarithmic radius
& $\xtil = \ln \left( \frac{R}{R_0} \right) $
& $\xtil = \ln \left( \frac{R}{R_0} \right) $
\\ 
$\Xtil$
& Scaled logarithmic radius
& $\Xtil = \ln \left( \frac{R}{\RH} \right) = \ln \Rtil $
& $\Xtil = \ln \left( \frac{R}{\RH} \right) = \ln \Rtil $
\\ 
$\sigutil$
& Radial velocity dispersion
& $\sigutil^2 = {1 \over {1+\gamma } }$
& $\sigutil^2 = {1 \over {1+\gamma+2\beta } }$
\\ 
$\gamma$
& Anisotropy parameter
& $\gamma= \frac{1}{\sigutil^2} - 1$
& $\gamma = \frac{1}{\sigutil^2} - 1 - 2\beta$
\end{tabular}

\setcounter{equation}{0}
\setcounter{figure}{0}
\setcounter{table}{0}
\appendix
\section{The Angular Momentum Function}\label{sec:FlmIntegration}
In this Appendix, we derive the forms of the angular momentum function
by explicitly performing the contour integration over angular momentum
$\Ltil$. Let us recall the angular momentum function is
\begin{equation}
\nonumber
	\Flm = { 1 \over {2\pi} } \int_0^\infty
	{ { e^{-i \eta {2\over {2-\beta}}  \ln \Ltil}} 
	\over { \lkapmom -\omtil \Ltil^{\twopm}} } 
\left[
	\left\{
	( \lkapmom ) \modterm - \gamma m \right\}
	 \Htil
	- m \Ltil \frac{d\Htil}{d\Ltil}
\right] { {d\Ltil}  \over \Ltil}.
\end{equation}
The integration is carried out over the variable $h$ defined by
\begin{xalignat}{2}
	h & = \twopm \ln \Ltil,
	& \quad
	\Htil (\Ltil) = \mathcal{H} (h).
\end{xalignat}
It is performed for three different cut-out factors, namely the
self-consistent disk, the inner cut-out disk and the doubly cut-out
disk (all defined in eq.~\eqref{eq:innercutoutfunction}).  We proceed
by splitting up the integral into two parts:
\begin{equation}
	\Flm(\eta) = \left\{
	( \lkapmom ) \modterm - \gamma m \right\} \flmint_1 - m \flmint_2,
        \end{equation}
where
\begin{xalignat}{2}
	\flmint_1 (\beta, \eta, \Ltilc)
	 & = { 1 \over {2\pi} } \twomp \int_{-\infty}^\infty
	{ { e^{-i \eta {2\over {2+\beta}}  h}} 
	\over { \lkapmom -\omtil e^h} } 
	\mathcal{H}(h) dh,
	\label{eq:J1}
&
	\flmint_2 (\beta, \eta, \Ltilc)
	& ={ 1 \over {2\pi} } \int_{-\infty}^\infty
	{ { e^{-i \eta {2\over {2+\beta}}  h}} 
	\over { \lkapmom -\omtil e^h} } 
	{{d\mathcal{H}} \over {dh} } dh.
\end{xalignat}
We need evaluate these only for the case $\beta=0$, since the general
integrals are related to the $\beta=0$ integrals as
\begin{xalignat}{3}
	\flmint_1 (\beta, \eta, \Ltilc) 
	& = \twomp \flmint_1 \left( 0, \hat{\eta}, \Ltilc^{\twopm} \right),
& 
	\flmint_2 (\beta, \eta, \Ltilc) 
	&= \flmint_2 \left( 0, \hat{\eta}, 	\Ltilc^{\twopm} \right),
& \text{with }
	\hat{\eta} & \equiv \eta \times \frac{2}{2+\beta}.
	\label{eq:hateta}
\end{xalignat}
$\Ltilc$ has been included as an argument, although clearly this is
only relevant to the doubly cut-out disk. We suppress the third
argument unless needed.

\subsection{The self-consistent disk}
\begin{figure}
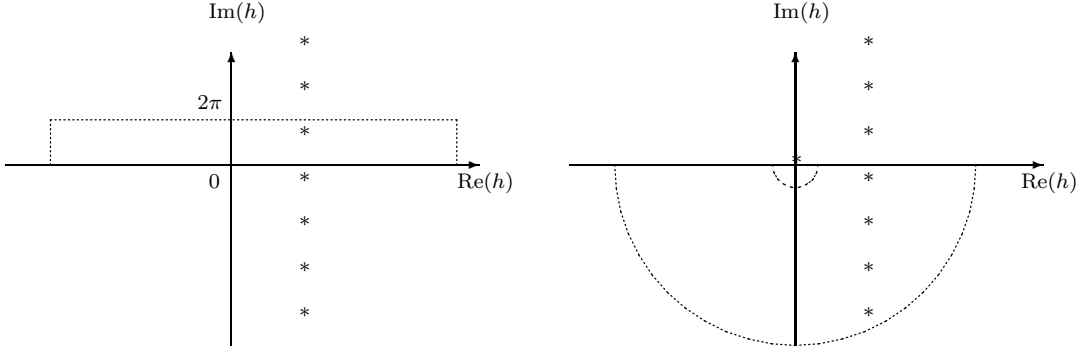

\setlength{\unitlength}{0.3mm}
\begin{picture}(300,180)(-20,100)


\put(0,200){\vector(1,0){210}}
\put(100,120){\vector(0,1){130}}
\put(90,265){Im($h$)}
\put(200,190){Re($h$)}

\dottedline{2}(20,200)(200,200)
\dottedline{2}(20,220)(200,220)

\dottedline{2}(20,200)(20,220)
\dottedline{2}(200,200)(200,220)

\put(85,225){$2\pi$}
\put(90,190){$0$}
\multiput(130,130)(0,20){7}{*}


\put(250,200){\vector(1,0){210}}
\put(350,120){\vector(0,1){130}}
\put(340,265){Im($h$)}
\put(450,190){Re($h$)}

\input{arc.tex}
\input{arcsmall.tex}

\put(348,197.5){*}

\multiput(380,130)(0,20){7}{*}

\end{picture} 
	\caption[Contours and poles for the
	self-consistent disk]{Contours and poles for the
	self-consistent disk. The left-hand plot shows the contour
	used for $\eta \ne 0$, and the right-hand that for $\eta =
	0$.\label{fig:rectcont1}}
\end{figure}

Here
\begin{xalignat}{2}
	\flmint_1 (0, \eta)
	 & = { 1 \over {2\pi} }  \int_{-\infty}^\infty
	{ { e^{-i \eta h}} 
	\over { \lkapmom -\omtil e^h} } 
	dh,
	& \quad
	\flmint_2 (0, \eta)
	&= 0.
\end{xalignat}
We first consider the case $\omega \ne 0$. $\flmint_1$ may be
evaluated by contour integration. The integrand has poles when $h_n =
\ln [(\lkapmom) / \omtil] + 2ni\pi$, where $n$ is an integer. The
poles occur at intervals along a line parallel to the imaginary
axis. Note that $\omtil$ is not in general real ($\omega = m\Omegap +
is$), so that the poles are displaced from the lines $h=2ni\pi$.
We evaluate the integral
\be
	\frac{1}{2\pi} \oint \frac{e^{\lambda h} e^{-i\eta h} }{ \lkapmom -\omtil e^h} dh,
\end{equation}
around a rectangular contour as shown in the left-hand plot of
fig.~\ref{fig:rectcont1}, with long sides at $h = 0$, $h=2i\pi$, and
short edges at $\pm L$, enclosing the pole at $h = \ln [ (\lkapmom) /
\omtil ]$. Here the logarithm denotes the principal value, i.e. the
imaginary part lies between $0$ and $2\pi$.
Then
\begin{equation}
\begin{split}
	\frac{1}{2\pi} & \int_{-L}^{L} \frac{e^{\lambda h}e^{-i\eta h} dh }{\lkapmom - \omtil e^h}
	-
	\frac{e^{2i\pi\lambda}e^{2\pi\eta}}{2\pi} \int_{-L}^{L} \frac{e^{\lambda h}e^{-i\eta h} dh }
	{\lkapmom - \omtil e^h}
	+
	\frac{1}{2\pi}  \int_{0}^{2\pi} \frac{ e^{\lambda L} e^{y \eta} e^{i\lambda y} e^{-i\eta L} dh }
	{\lkapmom - \omtil e^{iy} e^{L}}
\\	-
	\frac{1}{2\pi} &\int_{0}^{2\pi} \frac{ e^{-\lambda L} e^{y \eta} e^{i\lambda y} e^{i\eta L} dh }
	{\lkapmom - \omtil e^{iy} e^{-L}}
	=
	-\frac{i}{\lkapmom} e^{\lambda \ln \frac{\lkapmom}{\omtil} } 
	e^{-\eta\ln \frac{\lkapmom}{\omtil} } 
\end{split}
\end{equation}
Taking $L \rightarrow \infty$, and then $\lambda \rightarrow 0$, we obtain the result
\be
	\flmint_1 (0, \eta ) = - {  {i e^{-i \eta \ln { \lkapmom \over \omtil}} }
		\over { ( \lkapmom ) \left( 1- e^{2\pi\eta} \right)} }.
	\label{eq:J1Singetanonzero}
\end{equation}
Generalising this result to all $\beta$ as described above, we obtain
\begin{xalignat}{2}
	\Flm(\eta) &= - \twomp 
	\left( \modterm - \frac{\gamma m}{\lkapmom} \right)
	\frac
	{i e^{-i \hat{\eta} \ln { \lkapmom \over \omtil}} }
	{ 
	\left( 1- e^{2\pi\hat{\eta}} \right)} 
	\label{eq:FlmSingetanonzero},
& \quad
	\eta &\ne 0.
\end{xalignat}
This expression is singular when $\eta = 0$. As we now demonstrate,
$\Flm (\eta)$ contains a delta-function at $\eta=0$.  For the case
$\Omegap=0$, $s = 0$, we can integrate the expression for $\Flm
(\eta)$ directly:
\begin{equation}
	\flmint_1 (0, \eta)
	  =  \frac{1}{\lkapmom}
	{ 1 \over {2\pi} } \int_{-\infty}^\infty
	  e^{-i \eta   h} dh
	=  \frac{1}{\lkapmom} \delta(\eta).
\end{equation}
and so for the general power-law disk
\begin{equation}
	\Flm(\eta) 
	= \twomp 
	\left\{ \modterm - \frac{\gamma m}{\lkapmom} \right\} 
	\delta(\hat{\eta}).
\end{equation}
To obtain the expression for general $\Omegap$ and $s$, we write
\be
	\flmint_1 (0, \eta)
	=
	\flmint_1^{(1)} (0, \eta) +  \flmint_1^{(2)} \delta(\eta ),
	\label{eq:016}
\end{equation}
where $\flmint_1^{(1)} (0,\eta)$ is equal to our previous
expression~\eqref{eq:J1Singetanonzero} for $\flmint_1 (0,\eta)$.  We
integrate ~\eqref{eq:016} over $\eta$, thus smoothing out the
delta-function.
\be
	\int_{\eta=-\epsilon}^{+\epsilon} d\eta \, \flmint_1 (0,\eta)
	=
	\int_{\eta=-\epsilon}^{+\epsilon} d\eta \,  \flmint_1^{(1)} (0,\eta ) +  
	\flmint_1^{(2)} \int_{\eta=-\epsilon}^{+\epsilon} d\eta  \, \delta(\eta )
	\label{eq:011},
\end{equation}
We now let $\epsilon \rightarrow 0$. We see from the expression for
$\flmint_1^{(1)} (\eta)$ already obtained~\eqref{eq:J1Singetanonzero}
that $\flmint_1^{(1)} (\eta)$ is odd in the limit of vanishingly small
$\eta$. It therefore contributes nothing in~\eqref{eq:011}. Using the
expression~\eqref{eq:J1} for $\flmint_1 (0,\eta)$, we obtain
\begin{equation}
	\flmint_1^{(2)}
	=
	\lim_{\epsilon \rightarrow 0}
	  { 1 \over {2\pi} } 
	\int_{\eta=-\epsilon}^{+\epsilon} d\eta 
	\int_{h=-\infty}^\infty \, dh
	{ { e^{-i \eta  h}}  
	\over { \lkapmom -\omtil e^h} } 
\end{equation}
Swapping the order of integration and performing the integration over
$\eta$, we obtain
\be
	\flmint_1^{(2)}
	=
	\lim_{\epsilon \rightarrow 0}
	  { 1 \over {2\pi i} } 
	\int_{h=-\infty}^\infty \, \frac{dh}{h}
	\frac
	 { e^{i \epsilon  h} -  e^{-i \epsilon  h}}
	{ \lkapmom -\omtil e^h } 
\end{equation}
We evaluate this by carrying out two separate contour integrals.  For
the integral with $e^{-i \epsilon h}$, we close the contour in the
lower half-plane, as shown in the left-hand plot of
fig.~\ref{fig:rectcont1}.  As usual, we take the radius of the inner
semi-circle to zero, and the that of the outer semi-circle to
infinity. The integrand then vanishes on the outer semi-circle, and we
obtain
\begin{equation}
	  { 1 \over {2\pi i} } 
	\int_{h=-\infty}^{\infty} \, \frac{dh}{h}
	\frac{ e^{-i \epsilon  h} }{ \lkapmom -\omtil e^h } 
	=
	\frac{1}{\lkapmom}
	\sum_{n=-1}^{-\infty}
	\frac{ e^{-i \epsilon \ln\frac{\lkapmom}{\omtil} }
	e^{   2n\pi\epsilon} }
	{\ln\frac{\lkapmom}{\omtil} + 2in\pi }
	-
	 { 1 \over 2 } \frac{1 }{ \lkapmom -\omtil} 
\end{equation}
The integral with $e^{+i \epsilon h}$ is evaluated over an analogous
contour in the upper half-plane. We obtain
\begin{equation}
	  { 1 \over {2\pi i} } 
	\int_{h=-\infty}^{\infty} \, \frac{dh}{h}
	\frac{ e^{i \epsilon  h} }{ \lkapmom -\omtil e^h } 
	=
	-\frac{1}{\lkapmom}
	\sum_{n=0}^{\infty}
	\frac{ e^{i \epsilon \ln\frac{\lkapmom}{\omtil} }
	e^{   -2n\pi\epsilon} }
	{\ln\frac{\lkapmom}{\omtil} + 2in\pi }
	+
	 { 1 \over {2 } } 
	\frac{ 1 }{ \lkapmom -\omtil  } 
\end{equation}
Subtracting these two results and taking the limit $\epsilon
\rightarrow 0$, we obtain
\begin{equation}
	  { 1 \over {2\pi i} } 
	\int_{h=-\infty}^{\infty} \, \frac{dh}{h}
	\frac{e^{i \epsilon  h}- e^{-i \epsilon  h} }{ \lkapmom -\omtil e^h } 
	=
	\frac{1 }{ \lkapmom -\omtil} 
	+
	\frac{i}{\lkapmom}
	\sum_{n=-\infty}^{\infty}
	\frac{1 }{ 2n\pi - i\ln\frac{\lkapmom}{\omtil}  }
\end{equation}
The sum can be explicitly evaluated.
\begin{equation}
\begin{split}
	\sum_{n=-\infty}^{\infty} 
	\frac{1 }{2n\pi-iz}
&	=
	-\frac{1 }{iz}
	+
	\sum_{n=-1}^{-\infty} 
	\frac{1 }{2n\pi-iz}
	+
	\sum_{n=1}^{\infty} 
	\frac{1 }{2n\pi-iz}
	=
	\frac{i}{z}
	+
	2iz \sum_{n=1}^{\infty} 
	\frac{1}
	{(2n\pi)^2+z^2}
\\
&	=
	\frac{iz}{4\pi^2} 
	\sum_{n=-\infty}^{\infty} 
	\frac{1}{(n+iz/2\pi)(n-iz/2\pi)}
	=
	-\frac{1}{2 } \cot\frac{iz}{2}
	= 
	-\frac{i}{2 } \frac{e^{-z}+1}{e^{-z}-1}
\end{split}
\end{equation}	
where we have used the standard result~\cite[eq. (5.1.6.4)]{Wyn:Russian}
\be
	\sum_{n=-\infty}^{\infty} \frac{1}{(n+a)(n+b)}
	=
	\frac{\pi}{b-a} ( \cot\pi a - \cot \pi b)
\end{equation}
Thus we obtain
\be
	\flmint_1^{(2)}
	= \frac{1}{2(\lkapmom)}
\end{equation}
The angular momentum function for the self-consistent disk is then
\begin{equation}
	\Flm (\eta)
	=
	\twomp
	\left( \modterm - \frac{\gamma m}{\lkapmom}  \right)
\left(
	\frac{1}{2} \delta(\hat{\eta} )
	-
	\frac{ie^{-i \hat{\eta} \ln { \lkapmom \over \omtil}}}{1- e^{2\pi\hat{\eta}}}
\right).
	\label{eq:FlmSing}
\end{equation}
The corresponding result for the Toomre-Zang disk is (correcting a
typographical error in Zang (1976), eq. (Z3.43))
\begin{equation}
	\Flm (\eta)
	=
	\left( \gamma+1 - \frac{\gamma m}{\lkapmom}  \right)	
\left(
	\frac{1}{2} \delta(\eta )
	-
	\frac{ie^{-i \eta \ln { \lkapmom \over \omtil}}}{1- e^{2\pi\eta}}
\right).
\end{equation}

\subsection{The inner cut-out disk}
Here, we must evaluate
\begin{xalignat}{2}
	\flmint_1 (0, \eta)
	& = { 1 \over {2\pi} }  \int_{-\infty}^\infty
	{ { e^{-i \eta h}} 
	\over { \lkapmom -\omtil e^h} } 
	{ {e^{N h}} \over { e^{N h}  + 1} } dh,
&\quad
	\flmint_2 (0, \eta)
	&={ 1 \over {2\pi} } \int_{-\infty}^\infty
	{ { e^{-i \eta h}} 
	\over { \lkapmom -\omtil e^h} } 
	{ {N e^{N h}} \over { \left( e^{N h}  + 1 \right) ^2} } dh.
\end{xalignat}
Again the integrands have poles when $h = \ln [(\lkapmom) / \omtil] +
2ni\pi$. However, they now also have poles along the imaginary axis,
at $h_j = (2j-1)i\pi/N$, where $j$ is an integer.  We integrate around
the same rectangular contour as for the self-consistent disk
(fig.~\ref{fig:rectcont1}). As well as the pole at $\ln [(\lkapmom) /
\omtil]$, we now also enclose $N$ poles lying on the imaginary
axis. Adding up the residues from all these poles, we obtain the
result:
\begin{equation}
	\flmint_1 (0,\eta) = \frac{-i}{1- e^{2\pi\eta}} 
	\left\{
	  \frac{ ( \lkapmom )^{N-1} e^{-i \eta  \ln { \lkapmom \over \omtil}}}
	{( \lkapmom )^{N} + \omtil^{N} }
	- {1 \over {N}} \sum_{j=1}^{N}
	{\exprei \over {\lkapmom - \omtil \expimi} }
	\right\},
\end{equation}
\begin{equation}
\begin{split}
	\flmint_2 (0,\eta) =  - \frac{i}{1- e^{2\pi\eta}} \times
&\left\{
	 \frac{( \lkapmom )^{N-1}  N  \omtil^{N} e^{-i \eta  \ln { \lkapmom \over \omtil}}}
	{\left[( \lkapmom )^{N} + \omtil^{N} \right]^2}
\right.
\\
&\left.
+
	{1 \over {N}} \sum_{j=1}^{N}
	{\exprei \over {\lkapmom - \omtil \expimi} }
	\left[ i\eta - \frac{\omtil \expimi}{\lkapmom - \omtil \expimi }
	\right]
\right\}.
\end{split}
\end{equation}
For $\beta \ne 0$, we obtain the angular momentum function
\begin{equation}
\begin{split}
	\Flm(\eta \not= 0) & =  -  \frac{i}{1-e^{2\pi\hat{\eta}}}
 \times  \left\{
	\frac%
	{( \lkapmom )^{N-1} e^{-i \hat{\eta} \ln { \lkapmom \over \omtil}}} 
	{( \lkapmom )^{N} + \omtil^{N}}
\right.
\\
& \times
\left[
	\twomp 
\left( ( \lkapmom ) \modterm - \gamma m \right)
-
\frac{mN \omtil^{N}}
{( \lkapmom )^{N} + \omtil^{N}}
\right]
  \\  
&\hspace{-2cm} -
	{1 \over {N}} \sum_{j=1}^{N}
	{\expreid\over {\lkapmom - \omtil \expimi} }
\left.
\left[
	 \frac{ m \omtil \expimi}{\lkapmom - \omtil \expimi }
	 - i m \hat{\eta} 
+
	{ {2-\beta} \over {2+\beta} } 
	\left( ( \lkapmom ) \modterm - \gamma m \right) 
\right]
\right\}.
	\label{eq:FlmInneretanezero}
\end{split}
\end{equation}
We can use l'H\^{o}pital's rule to obtain the result for
$\eta=0$. This requires that both the numerator and denominator of
~\eqref{eq:FlmInneretanezero} are zero in the limit $\eta \rightarrow
0$.  The denominator $(1- e^{2\pi\eta})$ is obviously zero in this
limit, and it can be shown that the numerator is also (e.g. it is
readily apparent for $N=1$).  Application of l'H\^{o}pital's rule then
yields
\begin{equation}
\begin{split}
	\Flm(\eta=0) 
& =    - \frac{i}{2\pi} 
 \Biggl\{
	\frac%
	{i (\lkapmom)^{N-1} \ln { \lkapmom \over \omtil} } 
	{( \lkapmom )^{N} + \omtil^{N}}%
\Biggr. 
\left[
	\twomp 
\left( ( \lkapmom ) \modterm - \gamma m \right)
-\frac{mN \omtil^{N}}
{( \lkapmom )^{N} + \omtil^{N}}
\right]
  \\  
&	+
	{1 \over {N}} \sum_{j=1}^{N}
	\frac{1}{\lkapmom - \omtil \expimi} 
\\ & \times  
\Biggl(
	(2j-1)\frac{\pi}{N}
\bigr.
\bigl.
\left[
	 \frac{ m \omtil \expimi}{\lkapmom - \omtil \expimi }
+
	{ {2-\beta} \over {2+\beta} } 
	\left( ( \lkapmom ) \modterm - \gamma m \right)
\right]
	-im
\Biggr)
\Biggl.
\Biggr\}.
\end{split}
\end{equation}


\subsection{The doubly cut-out disk}

\noindent
We briefly summarise the results for the doubly cut-out disks. Here,
\begin{equation}
	\flmint_1 (0, \eta, \Ltilc)
	   = { 1 \over {2\pi} }  \int_{-\infty}^\infty
	{ { e^{-i \eta  h}} 
	\over { \lkapmom -\omtil e^h} } 
	{ { e^{N h} } \over { \left[e^{N h}  + 1 \right] 
	\left[\Ltilc^{-M_\beta} e^{M h} + 1 \right]} }
 dh,
\end{equation}
\begin{equation}
	\flmint_2 (0, \eta, \Ltilc)
	={ 1 \over {2\pi} } \int_{-\infty}^\infty
	{ { e^{-i \eta h}} 
	\over { \lkapmom -\omtil e^h} } 
	{ {N e^{N h} \left( \Ltilc^{-M_\beta} e^{M h} + 1 \right) 
	- M \Ltilc^{-M_\beta} e^{N h} e^{M h}}
	\over
	{\left[ e^{N h} +  1\right]^2 
	\left[ \Ltilc^{-M_\beta} e^{M h} + 1 \right] ^2} }
 dh.
\end{equation}
The integrand now has additional poles at $ h_k = (2k-1)i \pi /M +
(2+\beta)/(2-\beta) \times \ln\Ltilc$, where $k$ is an integer. We
thus obtain a second sum, and $\Flm$ becomes
\begin{equation}
\begin{split}
	\Flm(\eta \not= 0 ) 
&=    \frac{-i}{1-e^{2\pi\hat{\eta}}} \times
\left\{
	\frac%
	{
	( \lkapmom )^{N-1} \LtilcNo 
	\omtil^{M} 
	e^{-i \hat{\eta} \ln { \lkapmom \over \omtil} }
	} 
	{
	\left[ ( \lkapmom )^{N} + \omtil^{N} \right]
	\left[ ( \lkapmom )^{M} + \LtilcNo \omtil^{M} \right]
	} 
\right.
\\
& \hspace{0.5cm} \times
\left[
	\twomp 
	\left( ( \lkapmom ) \modterm - \gamma m \right)
	-\frac{mN \omtil^{N}}
	{( \lkapmom )^{N} + \omtil^{N} }
	+\frac{mM ( \lkapmom )^{M}}
	{( \lkapmom )^{M} + \LtilcNo \omtil^{M} }
\right]
\\   
&	-
	{ \LtilcNo \over {N}} \sum_{j=1}^{N}
	\frac{\expreid}
	{
	\left[ e^{ (2j-1) \frac{M}{N} i\pi } + \LtilcNo \right]
	\imbracki
	}
\\
&\hspace{0.5cm} \times
\left[
	 \frac{ m \omtil \expimi}{  \lkapmom - \omtil \expimi  }
	- i m \hat{\eta} 
+	\twomp \left( ( \lkapmom ) \modterm - \gamma m \right) 
\right]
\\
&	+
	{ {\Ltilc^{N_\beta}} \over {M}}
	e^{ -i \hat{\eta} \twopm \ln \Ltilc } 
	\sum_{k=1}^{M}
\frac
	{
	e^{ \frac{2k-1}{M} \pi \hat{\eta} } 
	}
	{
	\left[ \expmNiNo + \Ltilc^{N_\beta} \right]
	\imbracko
	}
\\
&\hspace{0.5cm} \times
\left.
\left[
	 \frac{ m \omtil  \expimo } { \lkapmom - \omtil \expimo  }
	- i m \hat{\eta} 
+	\twomp
	\left( ( \lkapmom ) \modterm - \gamma m \right) 
\right]
\right\}.
	\label{eq:FlmDoublyetanezero}
\end{split}
\end{equation}
In the limit $\eta \rightarrow 0$:
\begin{equation}
\begin{split}
	\Flm(\eta = 0 ) 
&=    -\frac{i}{2\pi} \times
\Biggl\{
	\frac%
	{
	i ( \lkapmom )^{N-1} \LtilcNo 
	\omtil^{M} 
	\ln { \lkapmom \over \omtil} 
	} 
	{
	\left[ ( \lkapmom )^{N} + \omtil^{N} \right]
	\left[ ( \lkapmom )^{M} + \LtilcNo \omtil^{M} \right]
	} 
\Biggr. \\  
&\hspace{0.5cm} \times
\left[
	\twomp 
	\left( ( \lkapmom ) \modterm - \gamma m \right)
	-\frac{mN \omtil^{N}}
	{ ( \lkapmom )^{N} + \omtil^{N} }
	+\frac{mM ( \lkapmom )^{M}}
	{ ( \lkapmom )^{M} + \LtilcNo \omtil^{M} }
\right]
\\   
&	+
	{ \LtilcNo \over {N}} \sum_{j=1}^{N}
	\frac{1}
	{
	\left[ e^{ (2j-1) \frac{M}{N} i\pi } + \LtilcNo \right]
	\imbracki
	}
\\
&\hspace{0.5cm} \times
\biggl[
	(2j-1) \frac{\pi}{N}
\biggl\{
	 \frac{ m \omtil \expimi}{ \left[ \lkapmom - \omtil \expimi \right] }
+	\twomp \left( ( \lkapmom ) \modterm - \gamma m \right) 
\biggr\}
	-im
\biggr]
\\
&	-
	{ {\Ltilc^{N_\beta}} \over {M}}
	\sum_{k=1}^{M}
	\frac{1}
	{
	\left[ \expmNiNo + \Ltilc^{N_\beta} \right]
	\imbracko
	}
\\
& \hspace{-2cm}\times
\biggl[
\left\{
	\frac{(2k-1)\pi}{M} - i \ln\Ltilc^{\twopm}
\right\}
\biggl\{
	 \frac{ m \omtil \expimo  } { \lkapmom - \omtil \expimo }
 \Biggr.
+	\twomp
	\left( ( \lkapmom ) \modterm - \gamma m \right) 
\biggr\}
-im
\biggr]
\Biggr\} .
\end{split}
\end{equation}

\end{document}